\documentclass[11pt, oneside]{article}   	% use "amsart" instead of "article" for AMSLaTeX format
\usepackage{geometry}                		% See geometry.pdf to learn the layout options. There are lots.
\usepackage{graphicx}				% Use pdf, png, jpg, or eps§ with pdflatex; use eps in DVI mode
\usepackage{hyperref}
\usepackage{amssymb}
\usepackage{tikz}
\usepackage{latexsym, amsmath, mathrsfs, graphicx}
\usepackage[utf8]{inputenc}
\newcommand{\beq}{\begin{eqnarray}}
\newcommand{\eeq}{\end{eqnarray}}
\usepackage{authblk}
\usepackage{xcolor}
\usepackage{esint}
\usepackage{empheq}
\usepackage{subcaption}
\usepackage[normalem]{ulem}
\usepackage{comment}

\usepackage[T1]{fontenc}
\usepackage[polish,english]{babel}
\usepackage[utf8]{inputenc}

\usepackage{accents}
\newcommand{\ubar}[1]{\underaccent{\bar}{#1}}

\graphicspath{{./figures/}}

\newcommand{\bfp}{\mathbf{p}}
\newcommand{\bfh}{\mathbf{h}}
\newcommand{\bfl}{\boldsymbol{\lambda}}
\newcommand{\bfm}{\boldsymbol{\mu}}
\newcommand{\bfbm}{\boldsymbol{\bar{\mu}}}

\newcommand{\umu}{\ubar{\mu}}

\newcommand{\oneph}{$1{\rm ph}$ }
\newcommand{\twoph}{$2{\rm ph}$ }
\newcommand{\twosp}{$2{\rm sp}$ }
\newcommand{\twosponeph}{$2{\rm sp}(1{\rm ph})$ }
\newcommand{\twosptwoph}{$2{\rm sp}(2{\rm ph})$ }

\newcommand{\ket}[1]{\vert{#1}\rangle}
\newcommand{\bra}[1]{\langle{#1}\vert}
\newcommand{\braket}[2]{\langle #1 \vert #2 \rangle}

\DeclareMathOperator*\Dprod{%
\mathchoice
  {\ooalign{$\bullet$\cr\hidewidth$\displaystyle\prod$\hidewidth\cr}}%
  {\ooalign{\scalebox{.7}{$\bullet$}\cr\hidewidth$\textstyle\prod$\hidewidth\cr}\mkern6mu}%
  {\mkern6mu\ooalign{\scalebox{.6}{$\bullet$}\cr\hidewidth$\scriptstyle\prod$\hidewidth\cr}\mkern3mu}%
  {\mkern4mu\ooalign{\scalebox{.5}{$\bullet$}\cr\hidewidth$\scriptscriptstyle\prod$\hidewidth\cr}\mkern2mu}}

\title{The relevant excitations for the one-body function in the Lieb-Liniger model}
\author{Mi{\l}osz Panfil and Felipe Taha Sant'Ana}
\affil{Faculty of Physics, University of Warsaw, ul. Pasteura 5, 02-093 Warsaw, Poland}
\date{\today}							% Activate to display a given date or no date

\begin{document}
\maketitle

\begin{abstract}
         We study the ground state one-body correlation function in the Lieb-Liniger model. In the spectral representation, correlations are built from contributions stemming from different excited states of the model. We aim to understand which excited states carry significant contributions, specifically focusing on the small energy-momentum part of the dynamic one-body function. We conjecture that relevant excitations take form similar to two-spinon states known from XXZ spin chain. We validate this hypothesis by numerical evaluation of the correlator with ABACUS algorithm and by analytical computations in the strongly interacting regime. 
\end{abstract}

\section{Introduction}

Investigations on dynamic correlation functions in the Lieb-Liniger model, and generally in quantum integrable models, have a long history. Recently, the interest was fuelled with new developments in experimental techniques with cold-atomic gases~\cite{Kinoshita:2006aa,Hofferberth:2007aa,Cazalilla_2011,Cheneau:2012aa,doi:10.1146/annurev-conmatphys-031214-014548} and theoretical developments in non-equilibrium dynamics~\cite{2008Natur.452..854R,2011RvMP...83..863P,2016JSMTE..06.4002E,2016JSMTE..06.4003C,1742-5468-2016-6-064006,2016JSMTE..06.4007V}. A standard approach to the problem is through the spectral sum over the form factors, matrix elements of the operators involved. Whereas our capabilities for an analytic evaluation of the full sum are limited to essentially free models, over the years techniques were developed to extract useful information about the correlation from partial summations. An example of this philosophy is a fully microscopic derivation of the Luttinger liquid predictions for ground state correlation functions in the Lieb-Liniger model and XXZ spin chain~\cite{1742-5468-2011-12-P12010,1742-5468-2012-09-P09001}.

Complementary approaches to the problem consist of explicit numerical evaluation of the sum. Here again, performing the sum over all eigenstates is not possible and therefore one has to scan the Hilbert space in search for important states. The ABACUS method~\cite{abacus} implements this approach and yields precise, experimentally relevant, predictions~\cite{Mourigal_2013,Lake_2013,PhysRevA.91.043617,PhysRevLett.115.085301}.

Recently, a new approach to dynamic correlation functions has originated within Generalized Hydrodynamics (GHD)~\cite{PhysRevX.6.041065,2016PhRvL.117t7201B,2017PhRvL.119v0604B}. A unified framework of GHD allows to address correlations in homogeneous and stationary states in the large-time and long-distance regimes~\cite{Doyon_correlations,Doyon_drude}. The most advanced, from that perspective, is the understanding of correlation functions of local conserved densities and currents, for which universal predictions can be formulated~\cite{doyon2020hydrodynamic,perfetto2020euler}. We refer to a recent review on the GHD approach of correlation functions and transport coefficients for more details~\cite{milosz_2021_2}.

In this work we follow the concept of thermodynamic form factors. Thermodynamic form factors are matrix elements between thermodynamically large states of a system with finite energy density. As such, they directly enter  the spectral sum in the thermodynamic limit and allow for computations of the dynamic correlation function in an arbitrary finite energy density state of the system. The thermodynamic form factors can be computed by either an appropriate limit of a finite system form factors~\cite{2018JSMTE..03.3102D} or through the thermodynamic bootstrap program~\cite{Bootstrap_JHEP,Cortes_Cubero_2020}. 

The understanding of relevant excitations is central to the thermodynamic form factors. In the case of particle-number conserving operators, like the local density, or other higher conserved local charges, these take form of particle-hole excitations~\cite{2018JSMTE..03.3102D}. The spectral sum is then organized in contributions from sectors with fixed number of particle-hole pairs. The leading low-energy and -momentum contributions come from one particle-hole pairs, while the next-to-leading come from two particle-hole pairs, and so on. Even in such circumstances, the complete evaluation of the spectral sum is challenging. Still, understanding the overall structure allowed for a number of interesting results. This includes generalization of the detailed balance~\cite{Foini_2017,De_Nardis_2017} and prediction of edge singularities~\cite{De_Nardis_2018} in non-equilibrium situations or inclusion of diffusion effects to the GHD equations~\cite{10.21468/SciPostPhys.6.4.049,2018PhRvL.121p0603D}. At the GHD level, it also led to an independent derivation of the Euler scale GHD equations~\cite{cubero2020generalized} and the verification of the GHD predictions for large-time and long-distance correlations in homogeneous systems~\cite{Cortes_Cubero_2020}. 

For other local operators, like creation and annihilation operators, the structure of the thermodynamic form factors is not clear. In this work we initiate studies in this direction, starting slowly with an attempt to understand relevant excitations for the zero-temperature, dynamic one-body function of the Lieb-Liniger model with a finite number of particles. We identify a class of excitations that are relevant for the low energy and momentum part of the correlation. These excitations have structure similar to two-spinon states known from XXZ spin chains. Although our findings concern ground-state correlations in a finite system, they are generalizable to finite energy density states in thermodynamically large systems. We will address this issue in a future work. 

The Lieb-Liniger model describes gases of bosonic particles confined to one spatial dimension~\cite{1998_Olshanii_PRL_81}. The particles interact through an ultra-local potential, which we consider to be repulsive. Also, we will only consider homogeneous systems where Bethe Ansatz methods can be directly applied. Systems in external potentials, however experimentally relevant, are more difficult to study with these methods. An exception is the strongly interacting regime, or the Tonks-Girardeau gas, where the exact solution in the presence of a trapping potential exists, allowing for analytical studies~\cite{2004JPhA...37.9335G,2018PhRvA..97c3609C,felipe}.

The classical results on the Lieb-Liniger model concern exact solution to the ground state~\cite{1963_Lieb_PR_130_1} as well as excitations above it~\cite{1963_Lieb_PR_130_2}, followed by its extension to the finite-temperature system at thermal equilibrium \cite{1969_Yang_JMP_10}. Since then the model has been extensively studied. In the following we recall some results concerning the one-body function.

In the Tonks-Girardeau regime, we note the results of Lenard for both the ground state~\cite{Lenard64} and finite temperature~\cite{Lenard66} of the static one-body function. These were later generalized to dynamic correlations in~\cite{one-body_TG_Korepin_Slavnov}, from which the asymptotics can be extracted~\cite{ITS1992351}. Finite interactions are much more difficult to study and there are only partial analytical results. Lenard's representation in terms of Fredholm determinants can be formally generalized to Fredholm minors with the help of auxilliary fields~\cite{one_body_Korepin_Slavnov,KorepinBOOK}. Alternatively, within the form factor approach, large-distance and long-time  asymptotes of the dynamic correlation one-body function could be derived~\cite{Karol2011_2, Karol2011_1}.

Regarding numerical approaches, notable results for the one-body correlation function at zero temperature come from the ABACUS method~\cite{1742-5468-2007-01-P01008} and Quantum Monte Carlo calculations \cite{Astrakharchik_1,Astrakharchik_2}, ranging from the strongly to the weakly interacting limits.

The Lieb-Liniger gas does not undergo Bose-Einstein condensation, such possibility is excluded by the HMW theorem~\cite{hohenberg,wagner}. However, a reminiscent of the condensation is the quasi-long range order found in the ground state correlations, and, related to it, small momentum divergence of the one-body function. The exact shape of this divergence can be predicted with the Luttinger liquid theory~\cite{Luttinger,Cazalilla_2004}. This phenomenological approach also predicts large-distance and long-time behaviour of the one-body function, in agreement with microscopic computations~\cite{Karol2011_1}.

The results presented in this paper bring in more intuitions about the relevant excitations behind the one-body function. The presentation is organized in the following way. In Sec.~\ref{BA} we summarize the basic concepts behind the Bethe Ansatz solution to the Lieb-Liniger model. We also introduce the one-body correlation function and the form factors which are the central quantities studied throughout this work. In Sec.~\ref{sec:spinons} we describe the elementary excitations above the ground state of the Lieb-Liniger model and introduce \emph{two-spinon states} as certain two-hole states. We also compute their dispersion relation. Then, in Sec. \ref{sec:abacus}, relying on the ABACUS calculations, we analyze the importance of the two-spinon states for the dynamic and static one-body function ranging from the weakly to the strongly interacting cases. In practice, we show that these excitations are responsible for small energy-momentum part of the correlator. Afterwards, we focus on the strongly interacting limit, or the Tonks-Girardeau gas. 
In Sec. \ref{sec:TG} we construct the form factors for the TG limit from the $(N+1)$-particle ground state with two-spinon and $m$ particle-hole excitations. Subsequently, we introduce an efficient ordering for the spectral sum such that contributions to it are strictly decreasing. We then explicitly compute the form factors for two classes of zero-momentum excitations, namely two-spinon and two-particle excitations to show that the former yields  larger contributions. Followed by that, we compare the importance of the two-spinon excitation and the particle-hole excitations over the one-body function, contrasting them with the exact result from Lenard's formula. 
In Sec. \ref{sec6} we work out the thermodynamic limit of the two-spinon contribution. Specifically, we show that two spinons carry significant contribution for systems composed of a thousand particles.
In Sec. \ref{summary} we summarize our results and mention possible further studies.
In App. \ref{appA} we evaluate the TG limit of the form factor and in App. \ref{appB} we provide a supplementary material for rewriting the spectral sum as integrals in the thermodynamic limit.

\section{Bethe Ansatz solution to the Lieb-Liniger model}\label{BA}

The Lieb-Liniger model describes a gas of bosonic particles interacting via a $\delta$-like potential. 
For a system of length $L$, its Hamiltonian, written in terms of canonical bosonic field operators, is~\cite{KorepinBOOK}
\begin{equation}
H = \int_0^L dx \left [\partial_x \Psi^\dagger (x)\partial_x \Psi(x) + c \Psi^\dagger (x)\Psi^\dagger (x)\Psi(x)\Psi (x) \right],
\end{equation}
in units where $\hbar=2m=1$\footnote{Moreover, we measure time in units of  $1/\epsilon_F$ and length in units of $1/k_F$, where $\epsilon_F = k_F^2$ and $k_F = \pi N/L$ are the Fermi energy and momentum.}. 
The operator of total number of particles, $\hat{N} = \int {\rm d}x \Psi(x)^\dagger \Psi(x)$, commutes with the Hamiltonian and the field-theory formulation in the subspace with fixed number of particles $N$ is equivalent to the Lieb-Liniger Hamiltonian~\cite{1963_Lieb_PR_130_1,1963_Lieb_PR_130_2}
\begin{equation}\label{eq.LL_I}
H = -\sum_{j=1}^N \partial_j^2 + 2c \sum_{j > l} \delta (x_j-x_l).
\end{equation}
The whole Fock space is then $\mathcal{H} = \otimes_{N} \mathcal{H}_N$ where $\mathcal{H}_N$ is the $N$-particle subspace. A unnormalized eigenstate $|\bfl\rangle$ of~\eqref{eq.LL_I} is described by a set of rapidities $\bfl = \{\lambda_j\}_{j=1}^N$. Upon imposing periodic boundary conditions, the rapidities are constrained through the Bethe equations 
\begin{equation}\label{eq.Bethe}
e^{i\lambda_j L} = \prod_{j \neq l} \frac{\lambda_j - \lambda_l + ic}{\lambda_j - \lambda_l -ic}, \quad  j = 1,\dots,N.
\end{equation}
For repulsive interactions, $c>0$, the solutions to the Bethe equations are real numbers. The set of obtained eigenstates is complete
\begin{equation}
	\hat{1} = \sum_{N} \sum_{\bfl \in \mathcal{H}_N} \frac{|\bfl\rangle \langle \bfl|}{\langle \bfl| \bfl\rangle}, \label{completeness}
\end{equation}
where the sum extends over all possible solutions to the Bethe equations~\eqref{eq.Bethe}. The energy and momentum of a given eigenstate are 
\begin{equation}
	E_{\bfl} = \sum_{j=1}^N \left(\lambda_j^2 - h\right) , \qquad P_{\bfl} = \sum_{j=1}^N  \lambda_j, \label{E_P}
\end{equation}
with the chemical potential $h$ controlling the density of the particles in the grand-canonical ensemble.
Furthermore, \eqref{eq.Bethe} can be reformulated in the logarithmic form
\begin{equation}\label{eq.Bethelog}
\lambda_j+\frac{1}{L} \sum_{l =1}^N \phi(\lambda_j - \lambda_l) = \frac{2\pi}{L}I_j, \quad  j = 1,\dots,N,
\end{equation}
where we identify the $I_j$'s as quantum numbers, that can be either integers or half-odd integers, depending on whether the number of particles $N$ is odd or even, respectively. The two-body phase shift $\phi(\lambda)$ is given by
\begin{equation}
	\phi(\lambda) = 2 {\rm arctan}(\lambda/c).
\end{equation}
The quantum numbers obey the Pauli principle, $I_j \neq I_l$ for $j \neq l$ --- otherwise the eigenstate is zero. Note also that the permutation of the quantum numbers leads to the same eigenstate. The sum in~\eqref{completeness} is carried out over proper sets of quantum numbers.  %Moreover, for an ordered set quantum numbers, the set of rapidities is also ordered, that is $\lambda_j > \lambda_l$ for $I_j > I_l$. 
The ground state of the model is given by the Fermi sea,
\begin{equation} \label{gs_qn}
	I_j^{\rm GS} = - \frac{N+1}{2} + j, \qquad j = 1, \dots, N
\end{equation} 
Simple excited states above the ground state are constructed by modifying one (or few) of the ground state quantum numbers. We expand this point in Section~\ref{sec:spinons}.

The normalization of the Hamiltonian eigenvectors $|\bfl\rangle$, in the Algebraic Bethe Ansatz formulation, is given by the Gaudin-Korepin formula~\cite{gaudin,korepin}. For a set of rapidities $\bfl$ that satisfy the Bethe equations we have
\begin{equation}
|| \bfl ||^2 \equiv \braket{\bfl}{\bfl} = c^N \prod_{j>k=1}^N \frac{\lambda_{jk}^2 + c^2}{\lambda_{jk}^2} \det_N \mathcal{G}(\bfl),
	\label{eq.norm}
\end{equation}
where the Gaudin matrix entries read 
\begin{equation}
	\mathcal{G}_{jk}(\bfl) = \delta_{jk} \left(L + \sum_{l=1}^N K (\lambda_j -  \lambda_l) \right) - K(\lambda_j - \lambda_k), \quad j, k = 1, \dots, N,
	\label{eq.gaudin}
\end{equation}
with
\begin{equation}\label{eq.kernel}
	K(\lambda) = \frac{\partial \phi(\lambda)}{\partial \lambda} = \frac{2c}{ \lambda^2 + c^2}.
\end{equation}

\subsection{One-body correlation function and the form factors}
Let us address the problem of calculating the one-body correlation function at zero temperature, 
\begin{equation}\label{eq.G(x,t)}
g(x,t) = \langle \bfl |\Psi^\dagger (x,t) \Psi (0,0)| \bfl\rangle,
\end{equation}
where $|\bfl\rangle$ is the ground state. Making use of the completeness property~\eqref{completeness} of the Bethe states, the correlation function, in the spectral representation, reads
\begin{equation}\label{eq.G(x,t)2}
g(x,t) = \sum_{\bfm \in \mathcal{H}_{N-1}} \frac{\bra{\bfl} \Psi^\dagger (x,t) \ket{\bfm} \bra{\bfm} \Psi(0,0) \ket{\bfl}}{\braket{\bfl}{\bfl}\braket{\bfm}{\bfm}},
\end{equation}
where the field operators $\Psi^{\dagger}(x)$ ($\Psi(0)$) increases (decrease) the number of particles by one.
Regarding that the observable $\Psi^\dagger(0)$ evolves in time and space according to 
\begin{equation}
	\Psi^{\dagger}(x,t) = e^{- i H t + i P x} \Psi^{\dagger}(0) e^{i H t - i P x},
\end{equation}
the one-body correlation function reads 
\begin{equation}
g(x,t) = \sum_{\bfm \in \mathcal{H}_{N-1}} e^{-i(E_{\bfl}-E_{\bfm})t + i(P_{\bfm}-P_{\bfl})x} |\langle \bfm | \Psi(0) | \bfl \rangle|^2,
\end{equation}
where the space- and time-independent term, $|\langle \bfm | \Psi(0) | \bfl \rangle|^2$, is the so-called form factor. The one-body function obeys a simple normalization condition
\begin{equation} \label{normalization}
	g(0,0) = n,
\end{equation}
where $n = N/L$ is the 1d gas density.

The field operator form factor for the Lieb-Liniger, computed by methods of Algebraic Bethe Ansatz and Quantum Inverse Scattering Method, is given by \cite{one_body_Korepin_Slavnov,1997CMaPh.188..657K,1742-5468-2007-01-P01008}
\begin{equation}
	|\langle \bfm | \Psi(0) | \bfl \rangle|^2 = c^{2N-1} \frac{\prod_{j>k=1}^N \left( \lambda_{jk}^2 + c^2\right)^2}{\prod_{j=1}^N \prod_{k=1}^{N-1} \left(\lambda_j - \mu_k \right)^2} \frac{\det_{N-1}^2 U(\bfm, \bfl)}{\| \bfm\|^2 \|\bfl\|^2}, \label{ff_LL}
\end{equation}
where we abbreviate $\lambda_{jk} \equiv\lambda_j - \lambda_k$. The entries of the  $(N-1)\times(N-1)$ matrix $U(\bfm, \bfl)$ are given by 
\begin{equation}\label{eq.U}
	U_{jk}(\bfm, \bfl) = \delta_{jk} \frac{V_j^+ - V_j^-}{i} + \frac{\prod_{a=1}^{N-1}(\mu_a - \lambda_j)}{\prod_{a \neq j}^N (\lambda_a - \lambda_j)} \left[ K(\lambda_j - \lambda_k) - K(\lambda_N - \lambda_k) \right], \quad j, k = 1, \dots, N-1,
\end{equation}
with
\begin{equation}
	V_j^{\pm} = \frac{\prod_{a=1}^{N-1} (\mu_a - \lambda_j \pm ic)}{\prod_{a=1}^N (\lambda_a - \lambda_j \pm ic)}.
	\label{eq.Vj}
\end{equation}

Our main interest will be the space-time Fourier transform of the one-body function defined as
\begin{equation}
	G(k, \omega) = \int_0^L {\rm d}x \int_{-\infty}^{\infty} {\rm d}t e^{i \omega t - i k x} g(x,t),
\end{equation}
which in the spectral representation becomes
\begin{equation} \label{G_komega}
	G(k,\omega) = 2\pi L \sum_{\bfm \in \mathcal{H}_{N-1}} \delta(\omega - E_{\mu} + E_{\lambda}) \delta_{k - P_{\mu} - P_{\lambda}}\, |\langle \bfm | \Psi(0) | \bfl \rangle|^2.
\end{equation}
Together with the normalization~\eqref{normalization}, it provides the sum rule
\begin{equation} \label{sum_rule}
	n = \sum_{\bfm \in \mathcal{H}_{N-1}} |\langle \bfm | \Psi(0) | \bfl \rangle|^2.
\end{equation}
We use such a sum rule to probe the effects of truncation over the spectral sum. In our considerations, we will also compute the static correlator and auto-correlation function defined as
\begin{equation} \label{G_k_omega}
	G(k) = \int_{-\infty}^{\infty} \frac{{\rm d}\omega}{2\pi} G(k, \omega), \qquad G(\omega) = \frac{1}{L} \sum_k G(k, \omega).
\end{equation}

\subsection{Tonks-Girardeau gas}

When the interparticle interaction becomes strongly repulsive, the system behaves similarly to the free Fermi gas, a phenomenon known as~\textit{fermionization}. Within such a regime, both the interacting bosonic system and the free fermionic one possess the same energy spectrum, whereas their wave functions are not the same. 
However, it is possible to described the bosonic many-body wave function in terms of the fermionic one~\cite{1960_Girardeau}. This correspondence holds also for reduced density matrices at both zero \cite{Lenard64} and finite temperatures \cite{Lenard66}. 

The strongly repulsive limit of the Lieb-Liniger model is the Tonks-Girardeau gas~\cite{1936_Tonks,1960_Girardeau}. In that case, the static one-body function, in its ground state, is~\cite{Lenard64,Zvonarev2005}
\begin{equation}\label{leny}
	g(x) = \det_N (I + V_1 +V_2) - \det_N(I+V_1),
\end{equation}
where $I$ is the identity matrix and 
\begin{equation}
	V_1^{ij} = -\frac{4}{L} \frac{\sin (x(k_i-k_j)/2)}{(k_i-k_j)} , \qquad V_2^{ij}=\frac{1}{L}e^{-ix(k_i+k_j)/2}, \qquad i,j = 1, \dots, N. 
\end{equation}

Lenard's formula also admits a thermodynamic limit in which the finite-size determinants become Fredholm determinants. Furthermore, there are generalizations to dynamic correlation functions~\cite{one-body_TG_Korepin_Slavnov} and to finite temperatures~\cite{Lenard66}. The latter, the static and finite-temperature case, involves a simple modification of the zero-temperature regime. This modification amounts to changing the underlying distribution of momenta from the zero-temperature ones to the finite-temperature Fermi-Dirac distribution. We will not go into more details because this work is solely focused on the ground-state correlations.

\section{Elementary excitations}\label{sec:spinons}
The excited states of the Lieb-Liniger model can be described in terms of elementary excitations, which are usually classified as particle (Lieb I) and hole (Lieb II) modes, with dispersion relations 
$\omega_{\pm}(k)$ \cite{1963_Lieb_PR_130_2}. They are created by adding a particle with a quantum number absent among the ground state quantum numbers or by removing one of the particles from the ground state, respectively. Such excitations change the number of particles. An arbitrary state of the Lieb-Liniger model can be described, in many equivalent ways, as a number of particle and hole excitations. 
Neutral excitations, in the sense of not changing the total particle number, can be created by combining both types into particle-hole pairs. 
Particle-hole excitations (ph in short) provide a natural and efficient organization of the spectral sum for two-point functions of operators that do not change the number of particles, 
see for example~\cite{2018JSMTE..03.3102D}. 
In this work we are interested in the one-body function, for which the spectral sum extends over states with one particle less than the ground state. 
For the $N$-particle correlation function, an obvious way to organize the spectral sum is then to understand the excited states as particle-hole excitations on top of the $(N-1)$-particle ground state.

\begin{figure}
\center
\includegraphics[scale=0.35]{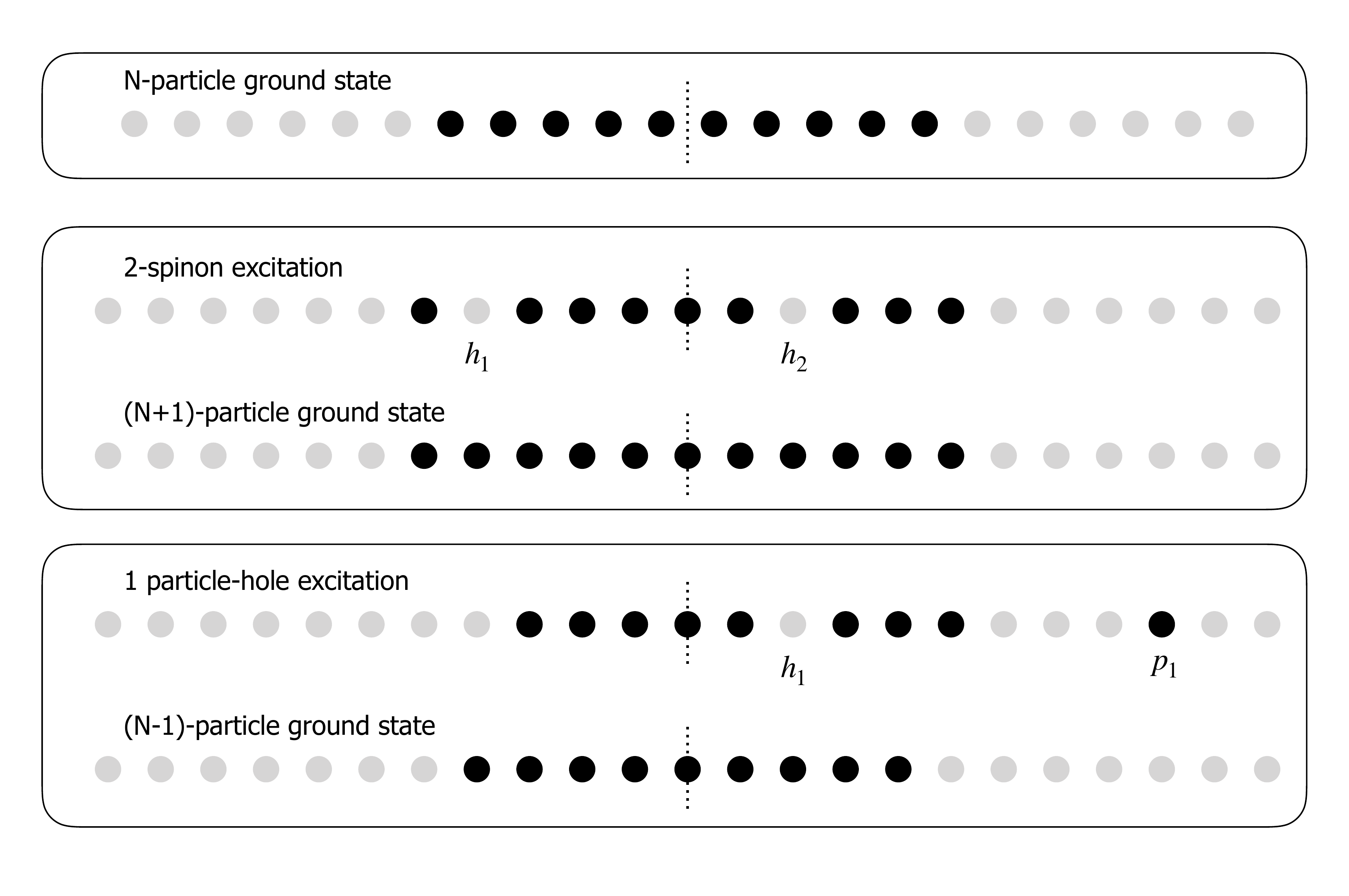}
\caption{(\emph{top}) Ground-state quantum numbers for $N=10$ particles. (\emph{middle}) A two-spinon excitation over the $(N-1)$-particle ground state. 
	It is best understood as a two-hole excitation over the $(N+1)$-particle ground state. 
	(\emph{bottom}) Standard one particle-hole excitation obtained from the $(N-1)$-particle 
	ground state by moving one particle from position $h_1$ to $p_1$.}
\label{fig:QN}
\end{figure}

Here instead, we propose an alternative way of spanning the spectral sum in which the excited states are created from the $(N+1)$-ground state. The basic excited states are then obtained by creating two-hole excitations. On top of them we can then add particle-hole excitations. Because of the formal analogy of the way the two-spinon states are described in the XXZ Hamiltonian (see for example~\cite{2017LNP...940.....F}) we refer to a ground state with two-hole excitations as a two-spinon state (\twosp in short)\footnote{Alternatively, we can think about the $2{\rm sp}$ states as a sum of three different classes of states: the ground state; one particle-hole ($1{\rm ph}$) excitation with the particle position 
fixed to the first empty slot on either side of the Fermi sea; or two particle-hole ($2{\rm ph}$) excitations with the particles positions fixed to the first empty slot on both sides of the Fermi sea.}. However, this analogy can only be taken that far. Unlike the actual spinon excitations in the spin chain that carry fractionalized spin~\cite{1981_Faddeev_PLA_85}, the spinon excitations that we introduce are not fractionalized excitations. 

We also emphasize that, independently of the way we organize the spectral sum, in the end the sum is over the same states. 
On the other hand, by centering the sum over excitations relevant for certain features of the correlator, we can hope to extract these features without performing the whole sum. 
This is the approach that we pursue here and we will show that the \twosp excitations capture the small energy-momentum part of the one-body function. 

To summarize, we propose the following organization of the spectral sum
\begin{equation}
	\sum_{\alpha \in \mathcal{H}_{N-1}} (\dots) = \sum_{m=0}^{\infty }\frac{1}{m! (m+2)!} \sum_{m{\rm ph}} \sum_{2{\rm sp}} (\dots).
\end{equation}
In this work we mainly focus on the $m=0$ contribution, showing that it saturates the small momentum and small energy part of the one-body function.

Two remarks are in place. Firstly, the ground-state correlation functions of the Lieb-Liniger model exhibit quasi-long range order, 
as phenomenologically captured by the Luttinger liquid theory~\cite{1981_Haldane_PRL_47,Cazalilla_2004,GiamarchiBOOK}. 
The power-law decay of the correlations is caused by accumulation of the excitations in the vicinity of the Fermi edges. 
Heuristically, we can understand them through the following picture. Creating an excitation, being it a particle, a hole, or a particle-hole pair, 
generates, within an interacting theory, a disturbance over other particles. In the case of the ground state, and due the the presence of the Pauli principle, 
most of particles are locked in the rapidities space. Only particles located at the edges of the Fermi sea possess relatively larger freedom. This leads, in large systems to a large number of small particle-hole excitations in the vicinity of the edges that, when properly accommodated for, are responsible for a long-range decay of the correlation functions~\cite{Karol2011_1,Karol2011_2}. Note that, despite the Tonks-Girardeau model being mappable to a free theory, its one-body function exhibits these generic features of an interacting theory (unlike the density fluctuations which are the same as for the free fermions). 
Similar phenomena also occur in higher energy parts of the spectrum and are responsible for edge singularities in the dynamic response functions~\cite{Imambekov_2008,Imambekov_2009,Kitanine_2012}. 

The lesson that we drew from this picture is that, for the correlations in quantum critical systems, it is not only enough to sum over relevant excitations, 
but one also needs to dress these excitations by the soft-modes --- small particle-hole excitations, see for example~\cite{1742-5468-2016-6-064006}. 
The soft-modes are subleading with respect to the actual (dispersive) excitations in the sense that their momentum and energy scales with the system size as $1/L$. 
Instead, the dispersive excitations carry finite momentum and energy in the thermodynamic limit. In practice, it means that when performing the spectral sum over the \twosp excitations, 
we also allow for small particle-hole excitations, denoted $(m{\rm ph})$, in the small window $\Delta J \ll N$ of quantum numbers in the vicinity of both Fermi edges. 
We set $\Delta J \approx \sqrt{N}$ such that $\Delta J/N \rightarrow 0$ in the thermodynamic limit. Moreover, we denote \twosp states dressed with \oneph and \twoph excitations as \twosponeph and \twosptwoph, respectively.

\begin{figure}
	\center
	\includegraphics[scale=0.5]{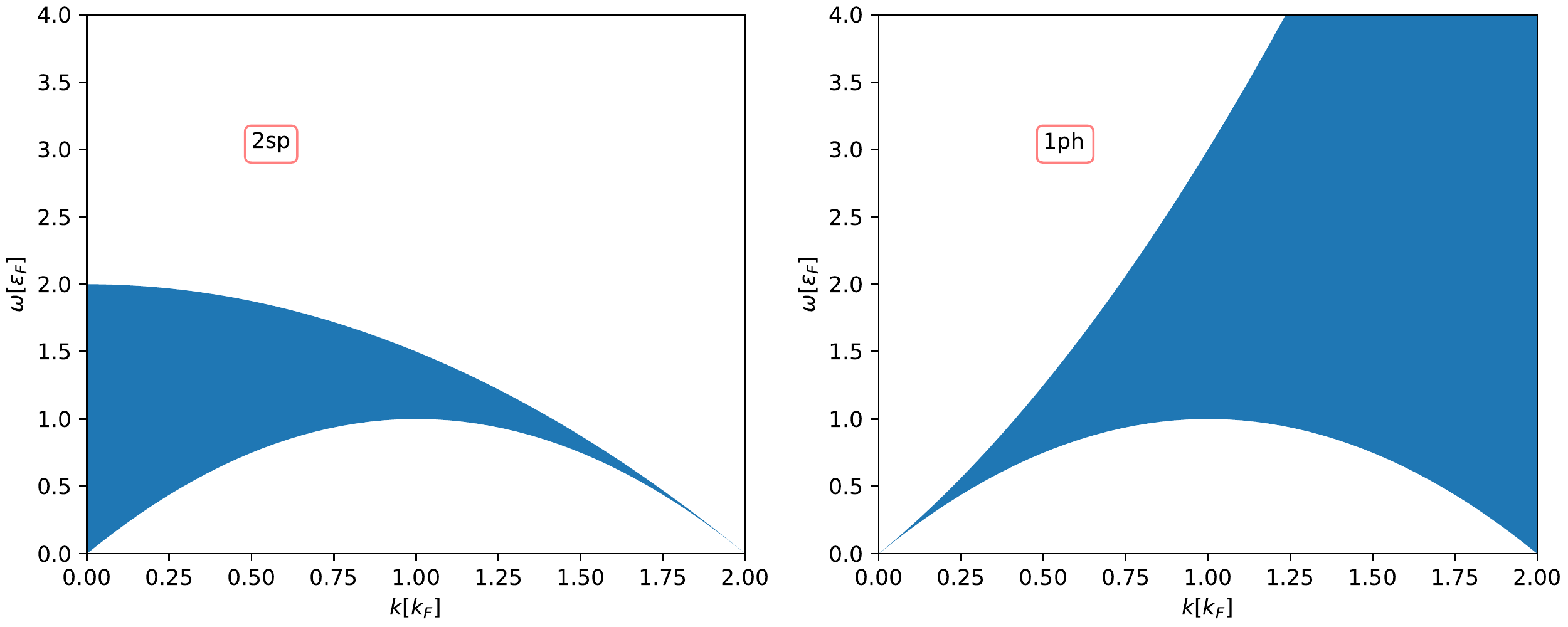}
	\caption{The continuum of 2sp (left) and 1ph (right) excitations in the Tonks-Girardeau gas. 
	The 2sp covers much better the low-momentum region of the spectrum. 
}
	\label{fig:dispersion}
\end{figure}

The second remark concerns the dispersion relation of the $2{\rm sp}$ excitations, which we will now compute. When we add or remove a particle from the system, the quantum numbers change from integers to half-odd integers or vice versa. 
However, when removing an even number of particles, the set of allowed quantum numbers does not change. This allows for a direct construction of the $2{\rm sp}$ state from $(N+1)$-particle ground state in which the only effect on the quantum numbers is the absence of the two chosen (removed) quantum numbers $I_1^-$ and $I_2^-$. The other quantum numbers remain intact, see fig.~\ref{fig:QN}. 
The situation is then analogous to the case of particle-hole excitations, in which only the quantum numbers related to the particle and to the hole are also modified. 
Recall that, in the thermodynamic limit, the momentum and energy of a particle-hole excitation with rapidities $\mu^+$ and $\mu^-$ are~\cite{KorepinBOOK}, 
\begin{align}
	k(\mu^+, \mu^-) = k(\mu^+) - k(\mu^-), \qquad
	\omega(\mu^+, \mu^-) = \varepsilon(\mu^+) - \varepsilon(\mu^-).
\end{align}
where the dressed momentum and energy are given by
\begin{align}
	k(\mu) = \mu + \int_{-q}^q {\rm d} \lambda\, F(\lambda|\mu), \qquad 
	\varepsilon(\mu) = \mu^2 - h + \int_{-q}^q {\rm d}\lambda\, (2\lambda) F(\lambda| \mu).
\end{align}
Here, the back-flow function $F(\lambda|\mu)$, is the solution of the following integral equation
\begin{equation}
	F(\lambda|\nu) = \frac{\theta(\lambda - \mu)}{2\pi} + \frac{1}{2\pi} \int_{-q}^q {\rm d}\mu K(\lambda - \mu) F(\mu| \nu).
\end{equation}
To complete the description, the Fermi rapidity $q$ is determined through the condition
\begin{equation}
	\int_{-q}^q {\rm d}\lambda\, \rho_{\rm p}(\lambda) = n.
\end{equation}
where $n$ is the 1d gas density and $\rho_{\rm p}(\lambda)$ obeys the Lieb equation~\cite{1963_Lieb_PR_130_2}
\begin{equation}
	\rho_{\rm p}(\lambda) = \frac{1}{2\pi} + \frac{1}{2\pi} \int_{-q}^q {\rm d}\mu\, K(\lambda - \mu) \rho_{\rm p}(\mu).
\end{equation}
As explained above, the two-hole case can be treated analogously to the case of a particle-hole excitation. Therefore, the momentum and energy of a state with two holes, $\mu_1^-$ and $\mu_2^-$, 
with respect to the $(N+1)$-particle ground state are $k(\mu_1^-, \mu_2^-) = - k(\mu_1^-) - k(\mu_2^-)$ and $\omega(\mu_1^-, \mu_2^-) = -\varepsilon(\mu_1^-) - \varepsilon(\mu_2^-)$, respectively. 
However, we view the two-spinon excitation as an excitation with respect to the $N$-particle ground state. 
Because of the incorporation of the chemical potential in~\eqref{E_P}, the formula does not change, \textit{i.e.}, 
\begin{equation}
	k_{2{\rm sp}}(\mu_1^-, \mu_2^-) = - k(\mu_1^-) - k(\mu_2^-), \qquad \omega_{2{\rm sp}}(\mu_1^-, \mu_2^-) = - \varepsilon(\mu_1^-) - \varepsilon(\mu_2^-).
\end{equation}
In fig.~\ref{fig:dispersion} we compare the range of momenta and energies covered by $2{\rm sp}$ and $1{\rm ph}$ excitations. 

In the Tonks-Girardeau gas this description becomes explicit. 
In such a limit, $\rho_{\rm p}(\lambda) = 1/(2\pi)$ and $q = \pi n$. The back-flow then disappears and the dressed momenta and energies are equal to the bare ones. 
Therefore,
\begin{equation}
	k_{2{\rm sp}}(\mu_1^-, \mu_2^-) = - \mu_1^- - \mu_2^-, \qquad \omega_{2{\rm sp}}(\mu_1^-, \mu_2^-) = 2\epsilon_F - (\mu_1^-)^2 - (\mu_2^-)^2, \label{k_w_2sp_TG}
\end{equation}
where we used the Fermi energy $\epsilon_F = k_F^2$ and the Fermi momentum $k_F = \pi n$.

\section{ABACUS results}\label{sec:abacus}
In this section, we perform an analysis on the excitations that contribute to the one-body function using the ABACUS method~\cite{abacus}. 
The ABACUS approach relies on a numerical evaluation of the spectral sum. The algorithm performs a scan through the Hilbert space resulting in a set of raw data comprising: 
the identification of the excited state, its momentum and energy, and the form factor. The raw data can be then summed up, using~\eqref{G_komega}, resulting in the dynamic correlation function $G(k, \omega)$. 
The quality of the result is then checked through the sum rule~\eqref{sum_rule}. 
In all present data, the sum rule is saturated to at least $99.8\%$. The first ABACUS computations of the one-body function in the Lieb-Liniger model were reported in~\cite{1742-5468-2007-01-P01008}.

\begin{figure}[h]
\center
\includegraphics[scale=0.6]{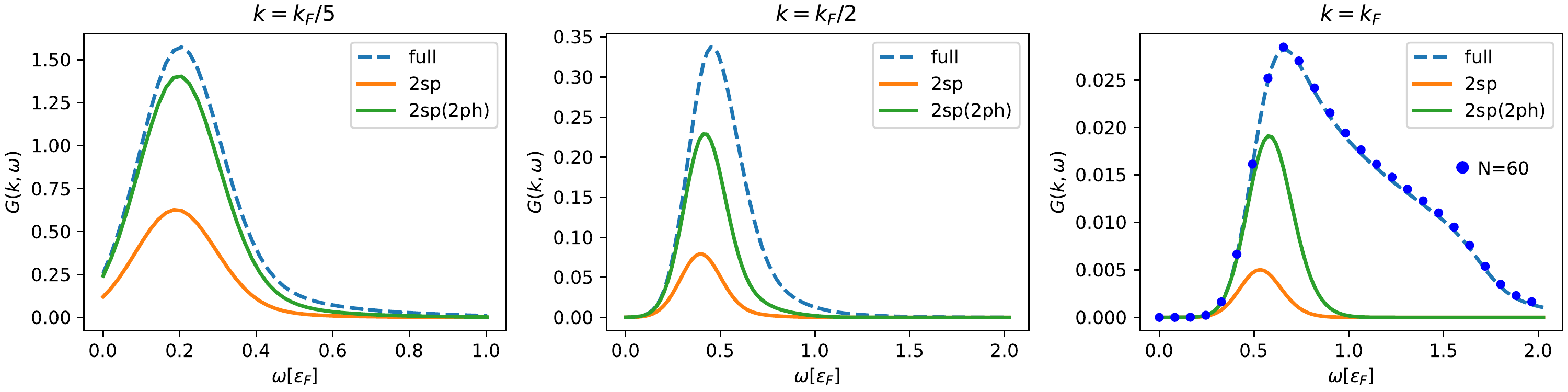}
\caption{Fixed momentum cuts of $G(k, \omega)$ for $c=4$, $N=100$ and for $k=k_F/5$, $k=k_F/2$ and $k = k_F$, from right to left. 
	We plot the 'full' ABACUS results together with the bare $2{\rm sp}$ and the dressed $2{\rm sp}(2{\rm ph})$ contributions. 
	We observe that the bare $2{\rm sp}$ excitations are not enough. Instead, they need to be dressed with particle-hole excitations in the vicinity of the Fermi edges. 
	For the system size considered, it is enough to include $2{\rm ph}$ excitations to saturate the correlation function at small energies even for momenta as high as $k_F$. For $k=k_F$  (third panel) we also show ABACUS results for $N=60$ to quantify the finite size effects.}
\label{fig:fixedK_ABACUS}
\end{figure}

We check the relevance of the $2{\rm sp}$ excitations in the following way. We consider the raw data produced by ABACUS and filter the excited states to include the 2sp states. 
As described in Sec.~\ref{sec:spinons}, we also allow for small particle-hole excitations within the distance $\Delta J$ (in quantum numbers) from both Fermi edges. 
In practice, we consider only one and two particle-hole excitations --- we observe that, increasing the number of particle-hole excitations, for the considered system sizes, yields marginal effects 
on the values of the correlators. Finally, we compare these results with the full ABACUS calculation.

In these numerical studies, we consider systems with unit density $N/L = 1$ containing $N=100$ particles at zero temperature. 
Also, we consider three values of the interaction parameter: $c = 1/4$, $4$ and $128$, which respectively correspond to regimes of weak, intermediate and strong interactions. 
We set $\Delta J = 5$, and this choice will be justified in Sec.~\ref{sec:TG}. 

\begin{figure}[h]
\center
\includegraphics[scale=0.6]{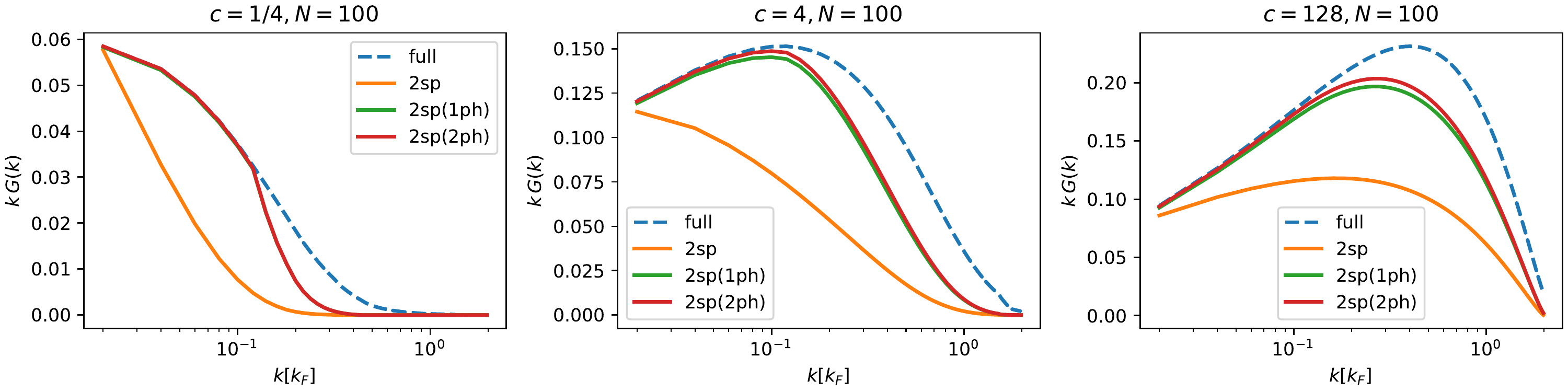}
\caption{Static correlator $G(k)$ \eqref{G_k_omega} for $c=1/4$, $c=4$ and $c=128$. We observe that the dressed $2{\rm sp}$ excitations saturate the correlation function at small momenta. 
	For $c=4$ and $c=128$, it is possible to observe how the inclusion of $2{\rm ph}$ excitations extends the approximation towards larger momenta.}
\label{fig:SSF_ABACUS}
\end{figure}

The dynamical one-body correlation function $G(k, \omega)$~\eqref{G_komega} as a function of $\omega$, for three different values of momentum $k$ and for $c=4$, is plotted in Fig.~\ref{fig:fixedK_ABACUS}. 
We observe that, even for relatively high momenta of order $k_F$, the low-energy tail is correctly reproduced with the dressed $2{\rm sp}$ excitations. 
For the interaction parameters $c=1/4$ and $c=128$, we qualitatively find the same results. 

\begin{figure}[h]
\center
\includegraphics[scale=0.6]{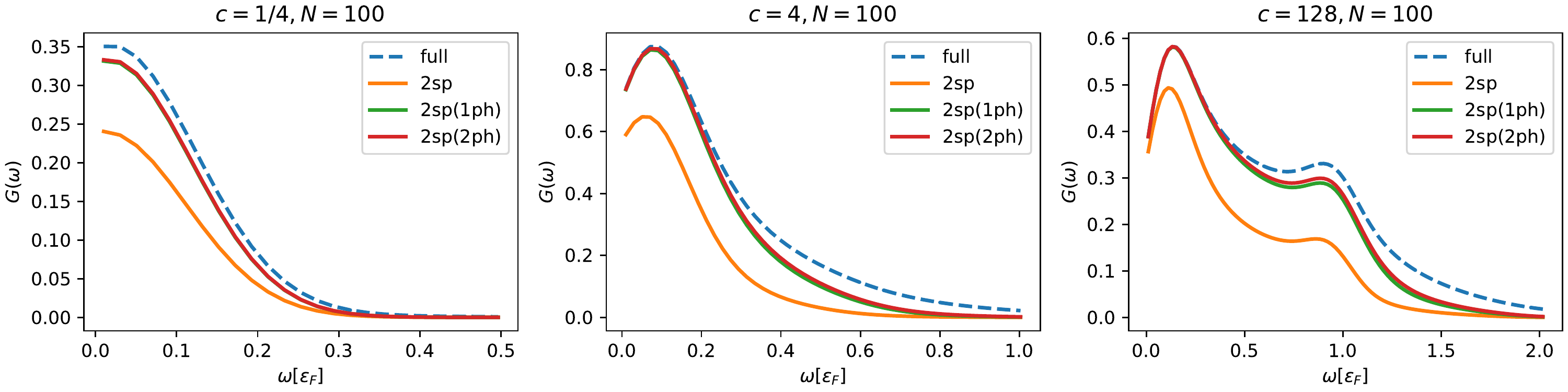}
\caption{Autocorrelation function $G(\omega)$~\eqref{G_k_omega} for the same three systems as in Fig.~\ref{fig:SSF_ABACUS}. The dressed $2{\rm sp}$ excitations saturate the correlator 
	at small energies (the discrepancy at $\omega \approx 0$ for $c = 1/4$ is caused by low-energy excitations at finite momentum that are present in a weakly interacting gas, see~\cite{1742-5468-2007-01-P01008}.) }
\label{fig:ASF_ABACUS}
\end{figure}

In Figs.~\ref{fig:SSF_ABACUS} and~\ref{fig:ASF_ABACUS}, we consider the integrated one-body functions, namely the static correlator $G(k)$ 
and the autocorrelation function $G(\omega)$ defined in~\eqref{G_k_omega}.
We observe that, also in the case of integrated one-body functions, their small momenta and energies regimes are very well described by the dressed \twosp excitations. 
This observation extends the validity of our approximation beyond the regime where both the momentum and the energy are small. Indeed, the static correlation function $G(k)$, in principle, 
has contributions from all energies. Similarly, the autocorrelation function $G(\omega)$ has contributions from all momenta. Only in the case of the autocorrelation function in the weakly interacting gas, 
first panel of fig.~\ref{fig:ASF_ABACUS}, we observe small contributions not captured by the dressed \twosp excitations.

\begin{table}[htp]
\begin{center}
\begin{tabular}{|l|r|r|r|r|}
	& \twosp & \twosponeph & \twosptwoph & full \\
	\hline 
	$c=1/4$ &  109 & 980 & 2521 & 3163 \\
	$c=4$ & 4516 & 68250 & 234850 & 806392 \\
	$c=128$ & 5050 & 186684 & 712355& 2108044 \\
	\hline
	total & 5050 & 451650 & 8793225 & ---
\end{tabular}
\end{center}
\caption{Number of states included in the spectral sum for the one-body function in the Lieb-Liniger gas. The last column shows the number of states in the raw ABACUS data, while the preceding ones 
	display the number of the corresponding excited states. The bottom row shows the total number of $2{\rm sp}$ together with dressings. 
	These numbers come from simply counting the number of $2{\rm sp}$ states in a system of $N=100$ particles together with dressing excitations in the window $\Delta J = 5$.}
\label{tab:abacus}
\end{table}

In table~\ref{tab:abacus}, we compare the number of states included in the spectral sum. The numbers show that the dressed $2{\rm sp}$ excitations compose a fraction of all states included with ABACUS. 
Yet, they are enough to capture the small momentum and energy sector of the one-body function. At the same time, the ABACUS algorithm heuristically scans over relevant excitations and, therefore, 
only a fraction of all dressed $2{\rm sp}$ excitations is considered. 

From table~\ref{tab:abacus} we also see that, from the point of view of the spectral sum, the strongly interacting gas is the most complicated regime, as it involves the largest number of contributing states. 
Therefore, in the remaining part of this work we focus on that limit. At the same time, the mathematical description of the Tonks-Girardeau gas is much simpler than the Lieb-Liniger model. 
This allows us to directly explore how the dressed two spinons contribute to the one-body function.

\section{Tonks-Girardeau gas}\label{sec:TG}

We specialize now to the Tonks-Girardeau gas, the $c\rightarrow \infty$ limit of the Lieb-Liniger model. We first compute the form factors for the dressed $2{\rm sp}$ excitations. 
Then, we use such a result to show that adding a new particle-hole leads to a state with a strictly smaller form factor. This allows us to organize the spectral sum in a way in which 
the values of the form factors monotonically decrease. Following that, we consider certain simple classes of zero-momentum excited states to show that the $2{\rm sp}$ states, among them, 
possess larger form factors than the zero-momentum states involving particle-hole excitations. Finally, we evaluate the static correlation function and compare it to Lenard's formula~\footnote{At this point we mention that some initial ideas for considering different classes of excited states in the computation of the one-body function were explored in the Master Thesis~\cite{Krzysztof_MT} of Krzysztof Dębowski prepared under the supervision of one of us.}.

To conclude this introductory part, we recall that, in the limit $c\rightarrow \infty$, the Bethe equations~\eqref{eq.Bethe} decouple to simple quantization conditions
\begin{equation}
\lambda_j =  \frac{2\pi}{L}I_j, \qquad j=1, \dots, N,
\end{equation}
with the Pauli principle still in play.  

\subsection{Form factors in the Tonks-Girardeau gas} \label{subsec:TG_ff}

We start by rewriting the form factor in a convenient way for the study of $2{\rm sp}$ states and particle-hole excitations above them. 
In App.~\ref{appA}, we compute the limit $c \rightarrow \infty$ of the form factor~\eqref{ff_LL}. The result is
\begin{equation}
	|\langle \bfm | \Psi(0) | \bfl \rangle|^2 = \frac{1}{2}\left(\frac{2}{L}\right)^{2N-1} \frac{\prod_{j > k =1}^N (\lambda_j - \lambda_k)^2  \prod_{j > k =1}^{N-1} (\mu_j - \mu_k)^2}{\prod_{j=1}^N \prod_{k=1}^{N-1} \left(\lambda_j - \mu_k \right)^2}. \label{ff_TG}
\end{equation} 
Given the product structure of the form factor, its format can be adjusted to the excitations that we consider. A generic excited state consists of two spinons parametrized by the positions of the two holes, $(h_1, h_2)$, and $m$ particle-hole excitations, each described by a pair $(p_j, h_j)$ with $j = 3, \dots, m+2$. Recall that such an excited state can be conveniently created from 
the $(N+1)$-particle ground state. Therefore, we can express the form factor as a form factor between the $N$-particle ground state and the $(N+1)$-particle ground state modified by the presence 
of the two spinons and $m$ particle-hole excitations. To this end, let us denote $\bfbm$ the set of ground state rapidities for $N+1$ particles. The rapidity set $\bfm$ for the excited state is then 
\begin{equation}
	\bfm = \bfbm - \bfh_{m+2} + \bfp_m.
\end{equation}
This notation implies that we respectively remove and add the sets $\bfh_{m+2}$ and $\bfp_m$ from the set $\bfbm$ (the indices denote the respective cardinality of a set). 
Then, the product between the rapidities $\bfm$ can be written as 
\begin{equation}
	\prod_{j=1}^{N-1} f(\mu_j) = \prod_{j=1}^{N+1} f(\bar{\mu}_j) \times \frac{\prod_{a=3}^{m+2} f(p_a)}{\prod_{a=1}^{m+2} f(h_a)}.
\end{equation}
In a similar fashion, the double product over distinct rapidities yields 
\begin{equation}
	\prod_{j \neq k}^{N-1} f(\mu_j, \mu_k) = \prod_{j \neq k}^{N+1} f(\bar{\mu}_j, \bar{\mu}_k) \times \frac{\prod_{a=3}^{m+2} \prod_{j=1}^{N+1} f^2(\bar{\mu}_j, p_a)}{\prod_{a=1}^{m+2} \Dprod_{j=1}^{N+1} f^2(\bar{\mu}_j, h_a)} \times \frac{\prod_{a > b=1}^{m+2} f^2(h_a, h_b)  \prod_{a> b=3}^{m+2} f^2(p_a, p_b) }{\prod_{a=1}^{m+2} \prod_{b=3}^{m+2} f^2(h_a, p_b)},
\end{equation}
for a symmetric function $f(\lambda, \mu) = f(\mu, \lambda)$. Note that we have introduced the following notation for the product excluding terms with equal arguments,
\begin{equation} \label{d_product}
	\Dprod_{j} f(\bar{\mu}_j, h) = \prod_{\substack{j\\ \bar{\mu}_j \neq h}} f(\bar{\mu}_j, h).
\end{equation}
With these two equalities, the form factor~\eqref{ff_TG} becomes
\begin{align}
	|\langle \bfm | \Psi(0) | \bfl \rangle|^2 &= \Omega(L, N) \times \prod_{a=1}^{m+2} \frac{\prod_{j=1}^N (\lambda_j -  h_a)^2}{\Dprod_{j=1}^{N+1} (\bar{\mu}_j - h_a)^2} \prod_{a=3}^{m+2} \frac{\prod_{j=1}^{N+1} (\bar{\mu}_j -  p_a)^2}{\prod_{j=1}^N (\lambda_j -  p_a)^2}\nonumber \\
	&\times \frac{\prod_{a > b=3}^{m+2} (h_a -  h_b)^2  \prod_{a> b=1}^{m+2} (p_a - p_b)^2 }{\prod_{a=1}^{m+2} \prod_{b=3}^{m+2} (h_a - p_b)^2},\label{ff_2sp}
\end{align}
where
\begin{equation}
	\Omega(N, L) \equiv  \frac{1}{2}\left(\frac{2}{L}\right)^{2N-1} \frac{\prod_{j > k =1}^N (\lambda_j - \lambda_k)^2  \prod_{j > k =1}^{N+1} (\bar{\mu}_j - \bar{\mu}_k)^2}{\prod_{j=1}^N \prod_{k=1}^{N+1} \left(\lambda_j - \bar{\mu}_k \right)^2} = \frac{L}{4} G^4(3/2) \left[ \frac{G(N+1) G(N+2)}{G^2(N+3/2)} \right]^2 
\label{OmegaNL}
\end{equation}
is a factor which is independent of the excited state and that provides an overall normalization to the correlation function. Here $G(x)$ is the Barnes G-function and the derivation is shown in App.~\ref{subsec.prefactor}.

\subsection{Organization of the spectral sum}
Now, let us describe how we organize the spectral sum. In the process, certain care is required to ensure that each excited state is considered only once. 
Thus, in order to guarantee this, we will introduce a specific ordering for the excited states. 
In the second part of this section we will show that form factors, in that ordering, are strictly decreasing as we increase the number of particle-hole excitations. 

\begin{figure}[h]
	\center
	\includegraphics[scale=0.4, trim={5cm 15cm 5cm 2cm}, clip]{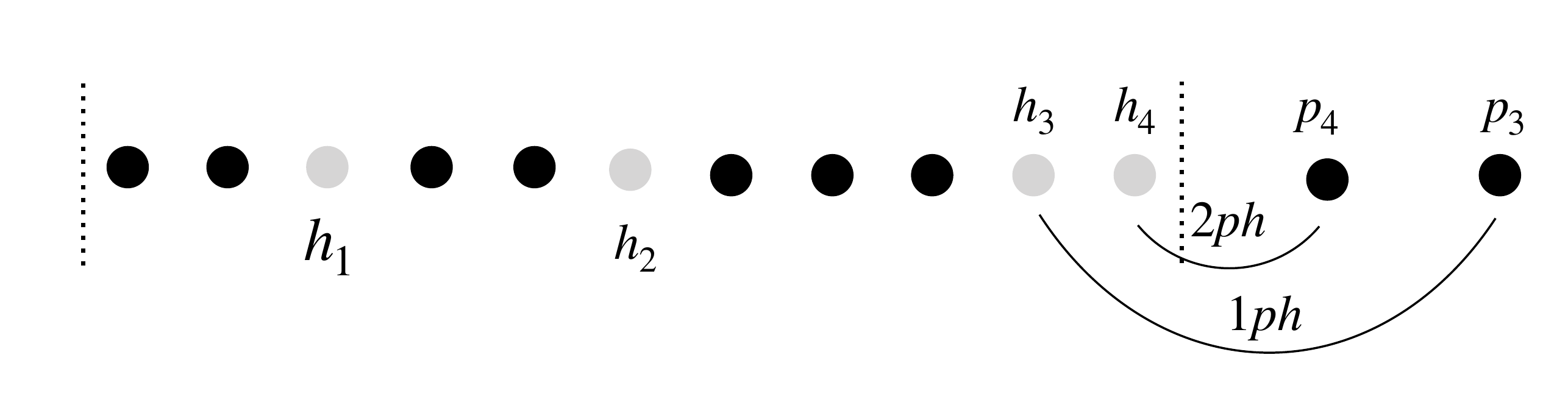}
	\caption{Example of the labelling of particles and holes in a 2sp(2ph) excited state.}
	\label{fig:labelling}
\end{figure}

Recall that the excited state, formed by a two-spinon and $m$ particle-hole excitations, is specified once we fix the positions of the $m+2$ holes and $m$ particles. 
We introduce the following prescription which allows us to uniquely label the excited state given $\bfp_m$ and $\bfh_{m+2}$. 
The procedure iteratively couples particles and holes into $m$ pairs. Then, the remaining two holes form the two-spinon excitation and are labelled such that $h_{1} < h_{2}$. 
In order to label the particle-hole pairs, we first search for the closest particle-hole pair. These are then labelled $p_{m+2}$ and $h_{m+2}$, respectively. 
The next closest pair is labelled $p_{m+1}$ and $h_{m+1}$, and so on. 
When there are two possibilities, we first choose a pair with negative momentum. In Fig.~\ref{fig:labelling} we show an example of the labelling of an excited state.

While spanning the spectral sum, we follow the ordering of the particle-hole excitations implied by the labelling. It means that whenever we add a new particle-hole excitation, the possible positions for the new particle and hole are such that the newly formed pair is the smallest present. As a consequence, the label of the new excited state simply appends the new particle-hole pair to the label of the original state. This way, from each excited state we have a manner of creating (a finite number of) descendent states. 
In the following, we show that the form factors of the descendent states are strictly smaller than the form factor of the parent state.

To begin with, let us consider the ratio between the form factors of a given excited state $|\bfm\rangle$ and the same state with an additional particle-hole excitation, $|\bfm + (p_{m+3}, h_{m+3})\rangle$. Such a ratio is then 
\begin{align}
	 \frac{|\langle \bfm + (p_{m+3}, h_{m+3}) | \Psi(0) | \bfl \rangle|^2}{|\langle \bfm | \Psi(0) | \bfl \rangle|^2} &= \frac{1}{(p_{m+3} - h_{m+3})^{2}}\,\frac{\prod_{j=1}^{N}(\lambda_{j} - h_{m+3})^{2}}{\Dprod_{j=1}^{N+1} (\bar{\mu}_j - h_{m+3})^2} \frac{\prod_{j=1}^{N+1}(\bar{\mu}_{j} - p_{m+3})^{2}}{\prod_{j=1}^{N}(\lambda_{j} - p_{m+3})^{2}} \nonumber \\
	&\times \frac{\prod_{b=1}^{m+2}(h_{m+3} - h_{b})^{2} \prod_{b=3}^{m+2}(p_{m+3} - p_{b})^{2}}{\prod_{b=1}^{m+2}(p_{m+3} - h_{b})^{2} \prod_{b=3}^{m+2}(h_{m+3} - p_{b})^{2}}. \label{ratio_ff}
\end{align}
We now proceed to proving that this ratio is smaller than one by showing that the factors in the first line and in the second line are independently smaller than one. 

Starting with the first term, let us parametrize the rapidities $p_{m+3}$ and $h_{m+3}$ with an index relative to the $(N+1)$-particle ground state
\begin{equation}
	p_{m+3} = \frac{2\pi}{L} \left( - \frac{N+2}{2} + P \right), \qquad h_{m+3} =  \frac{2\pi}{L} \left( - \frac{N+2}{2} + H \right),
\end{equation}
where $P \in \{\dots, -2, -1, 0, N+2, N+3, \dots\}$ and $H \in \{1, 2, \dots, N+1\}$. Then, we have that the first line of~\eqref{ratio_ff} becomes 
\begin{equation}
	\frac{1}{(P-H)^2} \frac{\prod_{j=1}^{N} (j-H+1/2)^2}{\Dprod_{j=1}^{N+1} (j-H)^2} \frac{\prod_{j=1}^{N+1} (j-P)^2}{\prod_{j=1}^{N} (j-P+1/2)^2} = \frac{1}{(P-H)^2}\frac{G_N(H)}{ G_N(P)}, %\equiv R_{\rm ff}(H,P),
\end{equation}
where we have defined
\begin{equation} \label{G_N}
G_N(x) =  \frac{\prod_{j=1}^{N} (j-x+1/2)^2}{\Dprod_{j=1}^{N+1} (j-x)^2}.
\end{equation}
Recall the prescription~\eqref{d_product} for the 'dotted product'. It amounts to skipping the terms that evaluate to zero. There always exists one such term for $x=H$. On the other hand, when $x=P$, there are no such terms and the product becomes an ordinary one.
The function $G_N(x)$ has the following property
\begin{equation}
	G_N(x) = G_N(N+2-x),%, \qquad B_N(P) = B_N( N+2 - P),
\end{equation}
reflecting the symmetry of the ground state. Moreover, for $x>0$, it can be represented through the Gamma function,
\begin{equation}
	G_N(x) = \frac{1}{\pi^2} \frac{\Gamma^2(x-1/2) \Gamma^2(N-x+3/2)}{\Gamma^2(x)\Gamma^2(N-x+2)}.%, \qquad B_N(P) =  \frac{\Gamma^2(P) \Gamma^2(P-N-1/2)}{\Gamma^2(P-1/2)\Gamma^2(P-N-1)}.
\end{equation}
Evaluating $G_N(x)$ for $x=H$,  we observe that it has maxima at the edges, that is for $x=1$ and $x=N+1$, given by 
\begin{equation}
	A = G_N(1) = G_N(N+1) = \frac{1}{\pi}\frac{\Gamma^2(N+1/2)}{\Gamma^2(N+1)}.
\end{equation}
The function $(P - H)^2 G_N(P)$, for any $P>N+2$, is bounded from above by choosing $H = N+1$. On the other hand, it is an increasing function of $P$. Therefore, we minimise it by choosing the edge value $P=N+2$, \begin{equation}
	B = G_N(0) = G_N(N+2) = \frac{4}{\pi} \frac{\Gamma^2(N+3/2)}{\Gamma^2(N+2)}.
\end{equation} 
Therefore, we have the following chain of relations
\begin{equation}
	\frac{1}{(P-H)^2}\frac{G_N(H)}{ G_N(P)} \leq \frac{A}{B} = \frac{1}{4} \left(\frac{N+1}{N+1/2} \right)^2 =  \left(\frac{1 + N}{1 + 2N} \right)^2 < 1,
\end{equation} 
which shows that the first line of~\eqref{ratio_ff} is indeed smaller than $1$.

From the construction, the new particle-hole pair is such that hole $h_{m+3}$ is the closest hole to the particle $p_{m+3}$. Then $|h_a - p_{m+3}| > |h_a - h_{m+3}|$. In a similar way, the new particle $p_{m+3}$ is the closest particle to the hole $h_{m+3}$, therefore $|p_a - h_{m+3}| > |p_a - p_{m+3}|$.
Using the two inequalities we readily obtain that 
\begin{equation}
	\frac{\prod_{b=1}^{m+2}(h_{m+3} - h_{b})^{2} \prod_{b=3}^{m+2}(p_{m+3} - p_{b})^{2}}{\prod_{b=1}^{m+2}(p_{m+3} - h_{b})^{2} \prod_{b=3}^{m+2}(h_{m+3} - p_{b})^{2}} < 1.
\end{equation}
This shows that the second line of~\eqref{ratio_ff} is also smaller than $1$. Therefore, the states in the spectral sum are organised such that the form factor of a descendent state is always smaller than the form factor of the parent state. 

\subsection{Two-spinon excited states versus particle-hole excited states} \label{subsec:versus_ph}

We now compare two ways of organising the spectral sum. First in terms of the two-spinon states with particle-hole excitations created from the $(N+1)$-particle ground state, and second in terms of the 'standard' particle-hole excitations created on top of the $(N-1)$-particle ground state. We perform the comparison on two levels. First microscopically, looking at values of specific form factors, second macroscopically by looking at the correlation function.

\begin{figure}[h!]
	\center
	\includegraphics[scale=0.5]{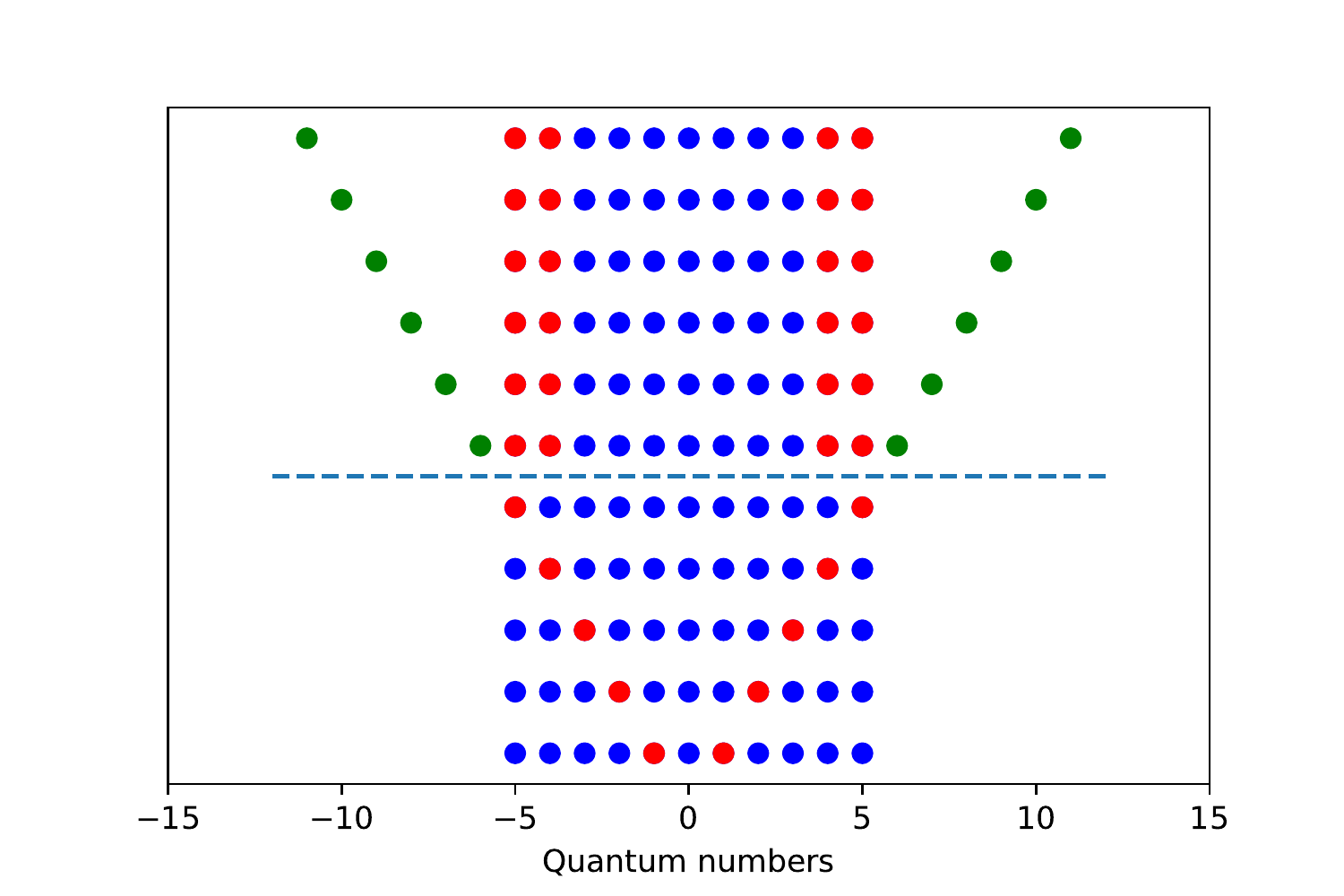}
	\caption{The two classes of zero-momentum excitations that we consider in Sec.~\ref{subsec:versus_ph}: the lower half is composed of two-spinon excitations and the upper half is composed of 2ph excitations with holes. The blue, red, and green colours correspond to, respectively, ground-state particles, hole excitations and particle excitations.}
	\label{fig:two_p_h_states}
\end{figure}

Let us start by considering two classes of zero-momentum excitations: a) formed by two spinons with opposite momenta and b) formed by two particle-hole excitations with opposite momenta created on top of the $(N-1)$-particle ground state with positions of holes at the Fermi edges. The two classes of excitations are shown in fig.~\ref{fig:two_p_h_states}.

\begin{figure}[h]
	\center
	\includegraphics[scale=0.5]{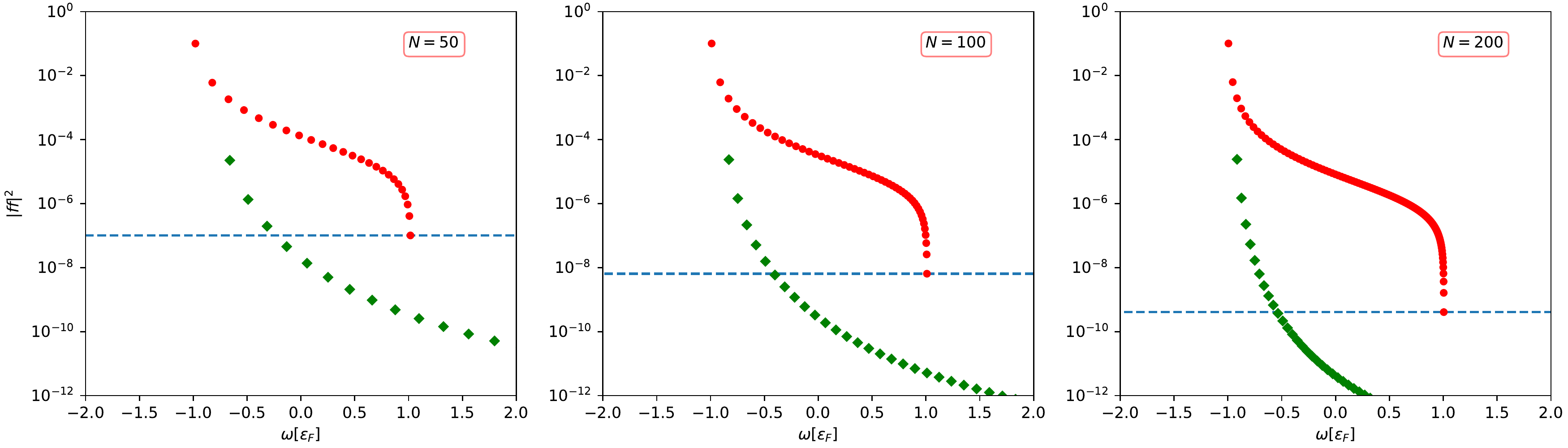}
	\caption{Form factors for the two classes of zero-momentum excitations depicted in Fig.~\ref{fig:two_p_h_states} for three different system sizes. We observe that form factors with two-spinon excitations are generally larger and the form factors for the particle-type excitations are only important for small excitations that we interpret as soft-modes dressings. The number $\Delta J$ can be estimated as $3, 5, 10$ for $N=50, 100$, and $200$ respectively. The dashed blue line highlights the smallest form factor with a two-spinon excitation.}
	\label{fig:two_p_h_ff}
\end{figure}

In Fig.~\ref{fig:two_p_h_ff} we show the results for the form factor. We conclude that the contribution from the two-spinon excitations is, on the average, a few orders of magnitude larger than from the two particles excitations. An exception is a handful of particle excitations localized in the vicinity of the Fermi edges. These excitations can be understood as dressings of the 2sp excitations. When performing the spectral sum we include these dressings by allowing for small particle-hole excitations in the vicinity of both Fermi edges.

These two families of zero-momentum states can be generalized to small momenta by breaking their parity symmetry. 
For example, for holes we could consider $h_1 = - h_2 +2\pi/L$.  
We have verified that the above conclusions also hold in such a case, \textit{i.e.}, the two-spinon excitations provide larger form factors and there is a relatively small number of particle excitations in the vicinity of the Fermi edges that yields comparable contribution.

In the second part of this section we compute the static one-body function by performing the spectral sum a) over dressed two-spinon excitations and b) over particle-hole excitations on top of the $(N-1)$-particle ground state. We have discussed above how the spectral sum over $2{\rm sp}$ states and their descendants is organized. The spectral sum over particle-hole pairs is simpler and can be implemented in the following way. The quantum numbers related to the holes belong to the finite set of the ground-state quantum numbers. On the other hand, the particles' quantum numbers are unbound. In practice, we introduce a cutoff quantum number $I_{\rm max}$ such that for any excited state $|I_j| \leq I_{\rm max}$. 
The number of potential excited states is then finite. Instead of talking about the cutoff quantum number, we introduce $N_{\rm max}^{\rm ph}$ as a number of possible slots for the newly created particles in the vicinity of the left and right Fermi edges. $N_{\rm max}^{\rm ph}$ also controls the number of possible slots for the holes in the vicinity of both Fermi edges. Therefore, the number of 1ph states is simply given by $(2N_{\rm max}^{\rm ph})^2$ and the number of 2ph states is $(2N_{\rm max}^{\rm ph}(2N_{\rm max}^{\rm ph}-1))^2/4$.

\begin{figure}
	\center
	\includegraphics[scale=0.5]{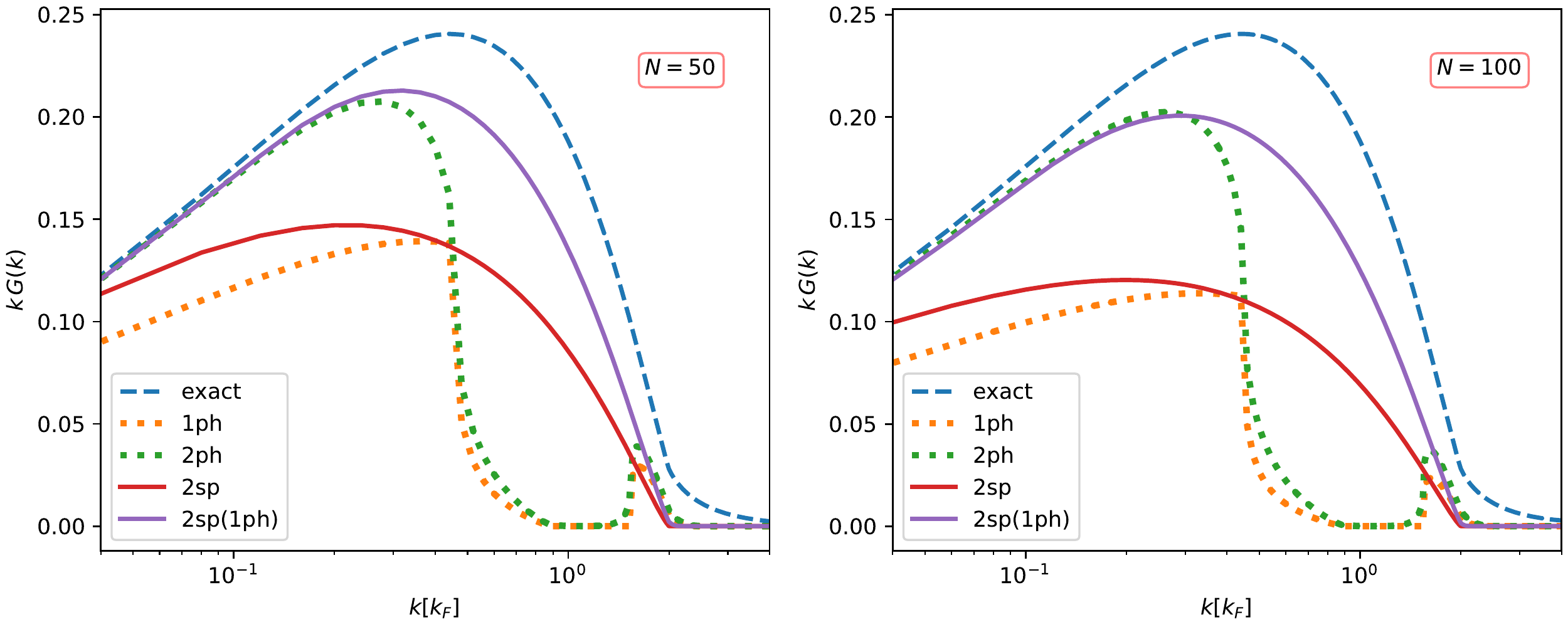}
	\caption{Plots of the static correlator for two ways of organizing the spectral sum: in terms of dressed two-spinon excitations and in terms of particle-hole excitations. We also plot the exact result~\eqref{leny} of Lenard, Fourier transformed to the momentum space.}
	\label{fig:SSF_TG_vs_ph}
\end{figure}

We now choose $N_{\rm max}^{\rm ph}$ such that the number of 2ph excited states is at least the number of \twosponeph states considered before, see table~\ref{tab:TG}. For $N = 50$ and with $N_{\rm max}^{\rm ph} = 11$ there are $53845$ states. For $N=100$ we simply double $N_{\rm max}^{\rm ph}$, which results in $896852$ states. In Fig.~\ref{fig:SSF_TG_vs_ph} we plot the resulting static correlator and compare it with the exact results and with the two-spinon approximation. The results clearly show that with smaller number of two-spinon excitations we can get much better coverage of the small momentum part of $G(k)$. 

\begin{table}[h!]
\begin{center}
\begin{tabular}{|l|r|r|r|r|}
& 2sp & 2sp(1ph) & 2ph \\
\hline 
$N=50$ & 1275 & 39810 &  53845\\
$N=100$ &  5050 & 451650 & 896852
\end{tabular}
\end{center}
\caption{Number of states included in the spectral sum for the one-body function in the Tonks-Girardeau gas in both the two-spinon and particle-hole approaches.}
\label{tab:TG}
\end{table}%

\section{Towards resummation of the form factors}\label{sec6}

In this final part we consider the thermodynamic limit and analytically perform the spectral sum over $2{\rm sp}$ states in the Tonks-Girardeau gas. Before proceeding with the computations, we note that any such partial contribution to the one-body function is vanishing in the thermodynamic limit and to obtain a finite contribution a summation over the soft-modes is required. We do not attempt it here. Instead, our aim here is to investigate how quickly the two-spinon contribution vanishes - we find that it does so in a relatively slow fashion as $(\log N)^2/\sqrt{N}$. This leads then to another question: how good an approximation is the two-spinon contribution for systems composed of hundreds particles? For a change we consider in this section the real space correlator.

We consider the thermodynamic limit of the spectral sum over two-spinon excitations,
\begin{equation}
	g_{2{\rm sp}}(x,t) = \sum_{|\alpha\rangle \in \mathcal{H}_{2sp}} e^{i \omega_{2{\rm sp}}(\alpha) t - i k_{2{\rm sp}}(\alpha) x} |\langle \alpha | \Psi(0) |{\rm GS}\rangle|^2.
\end{equation}
For the two-spinon state, the form factor~\eqref{ff_2sp} is   
\begin{align}
	|\langle \bfm | \Psi(0) | \bfl \rangle|^2 = \Omega(L, N) \times \frac{\prod_{j=1}^N (\lambda_j -  h_1)^2}{\Dprod_{j=1}^{N+1} (\bar{\mu}_j - h_1)^2}  \frac{\prod_{j=1}^N (\lambda_j -  h_2)^2}{\Dprod_{j=1}^{N+1} (\bar{\mu}_j - h_2)^2} (h_1 -  h_2)^2, 
\end{align}
and the sum extends over possible choices of the two holes $(h_1, h_2)$ in the $(N+1)$-particle ground state. The rapidities $\lambda_j$ and $\bar{\mu}_j$ are respectively ground-state rapidities of systems with $N$ and $N+1$ particles.
Recall the energy and momentum of the 2sp excitation from~\eqref{k_w_2sp_TG},
\begin{equation}
	\omega_{2{\rm sp}}(h_1, h_2) = 2\epsilon_F - h_1^2 - h_2^2, \qquad k_{2{\rm sp}}(h_1, h_2) = - h_1 - h_2.
\end{equation}
%\begin{align}
%	\lambda_j &= \frac{2\pi}{L} \left( - \frac{N+1}{2} + j\right), \quad  j=1, \dots, N, \\
%	\bar{\mu}_j &= \frac{2\pi}{L} \left(- \frac{N+2}{2} + j \right), \quad j = 1, \dots, N+1,
%\end{align}
We parametrize positions of the two holes by writing
\begin{equation}
	h_a = \frac{2\pi}{L} \left(- \frac{N + 2}{2} + k_a \right).
\end{equation}
The form factor reads then
\begin{equation}
	|\langle \bfm | \Psi(0) | \bfl \rangle|^2 = \left( \frac{2\pi}{L}\right)^2 \Omega(L, N)G_N(k_1) G_N(k_2) (k_1 -  k_2)^2,
\end{equation}
with $G_N(x)$ defined in~\eqref{G_N}. The spectral sum becomes
\begin{equation}
	g_{2{\rm sp}}(x,t) =  \frac{1}{2}\left( 2\pi n \right)^2  \Omega(L, N) \sum_{k_1, k_2=1}^{N+1} e^{i d(x,t; k_1/N)} e^{i d(x,t; k_2/N)} G_N(k_1) G_N(k_2) \left(\frac{k_1 - k_2}{N}\right)^2, \label{foo}
\end{equation}
where the factor $1/2$ takes into account that each 2sp state appears twice in the summation and
\begin{equation}
	d(x,t;k/N) = \left(\epsilon_F - \left(\frac{2\pi}{L}\right)^2 \left(-\frac{N+2}{2} + k\right)^2 \right)t + \frac{2\pi}{L} \left(- \frac{N+2}{2} + k\right) x.
\end{equation}
In practice, we can approximate $d(x,t;k/N)$ by writing 
\begin{equation}
	d(x,t;k/N) = 4 \epsilon_F t \left(\frac{1}{4} -  \left( \frac{k}{N} - \frac{1}{2} \right)^2\right) + 2 k_F x \left( \frac{k}{N} - \frac{1}{2} \right) + \mathcal{O}(1/N).
\end{equation}
The double sum over $k_1$ and $k_2$ in~\eqref{foo} can be factorized into products of single sums. We have
\begin{align}
	g_{2{\rm sp}}(x,t) =  \frac{\left( 2\pi n \right)^2 \Omega(L, N)}{\pi^4 N^2} \left( I_2(x,t) I_0(x,t) - I_1^2(x,t) \right). \label{foo2}
\end{align}
where
\begin{equation}
	I_a(x,t) = \pi^2 N \sum_{k}^{N+1} e^{i d(x,t; k/N)} G_N(k) \left(\frac{k}{N}\right)^a, \quad a = 0,1,2,
\end{equation}
and we introduced factor $\pi^2 N$ for future convenience.
The function $G_N(k)$ evaluated at $k = x (N+1)$, and in the limit of large $N$, is
\begin{equation}
	G_N(x (N+1)) \sim \frac{1}{\pi^2 N^2} \frac{1}{x(1-x)}, \quad N \rightarrow \infty.
\end{equation}
It diverges when $x$ approaches $0$ or $1$, which corresponds to $k$ in the vicinity of $1$ and $N+1$. In other words, when a hole approaches the edges of the Fermi sea. This divergence has then an effect on the sum and requires regularization in the continuum limit. In Appendix~\ref{appB} we show that
\begin{equation}
	\pi^2 N \sum_{k=1}^{N+1} G_N(k) f(k/N) =\fint_0^1 {\rm d}z \frac{f(z)}{z(1-z)} + \Xi_N (f(0) + f(1)) + \mathcal{O}(k^*/N^2), \label{single_sum}
\end{equation}
where $\Xi_N = \ln(N+1) + A + \gamma$ and the principal value integrals are defined in \eqref{pvint}. 
Using formula~\eqref{single_sum} for $I_a(x,t)$ with $a=0,1,2$, we find
\begin{align}
	I_a(x,t) &= \left( J_a(x,t)  + \Xi_N (e^{id(x, t; 1)} +  \delta_{a,0} e^{id(x, t; 0)})\right), \label{foo3}
\end{align}
with
\begin{equation}
	J_a(x,t) = \fint_0^1 {\rm d}z \frac{e^{i d(x,t; z)}}{(1-z)} z^{a-1}.
\end{equation}
We rewrite now the expression~\eqref{foo2} for $g_{2{\rm sp}}(x,t)$ in the large system size using formula~\eqref{foo3}. 
In doing so it is convenient to group different terms according to powers of $\Xi_N$. We then have
\begin{equation}
	g_{2{\rm sp}}(x,t) =  \frac{\left( 2\pi n \right)^2 \Omega(L, N)}{\pi^4 N^2} \left( \Xi_N^2 + \Xi_N K(x, t)  +  L(x, t) \right), \label{final_g_2sp}
\end{equation}
where
\begin{align}
	K(x, t) &= e^{id(x,t, 1)} \left(J_0(x,t) + J_2(x,t) - 2J_1(x,t) \right) + e^{id(x,t, 0)} J_2(x,t),\\
	L(x,t) &= J_0(x,t) J_2(x,t) - J_1^2(x,t).
\end{align}
This is the final expression for the $2{\rm sp}$ contribution to the dynamic one-body function in the large system at finite density. As predicted, the contribution vanishes in the thermodynamic limit, the exact scaling is $(\log N)^2/\sqrt{N}$ which is relatively slow (see App.~\ref{subsec.prefactor} for the asymptotic behavior of $\Omega(L, N)$). Therefore we can expect that for relatively large numbers of particles the contribution from $2{\rm sp}$ states remain significant. We can quantify this expectation by computing the sum rule~\eqref{sum_rule}.

\begin{figure}
	\center
	\includegraphics[scale=0.5]{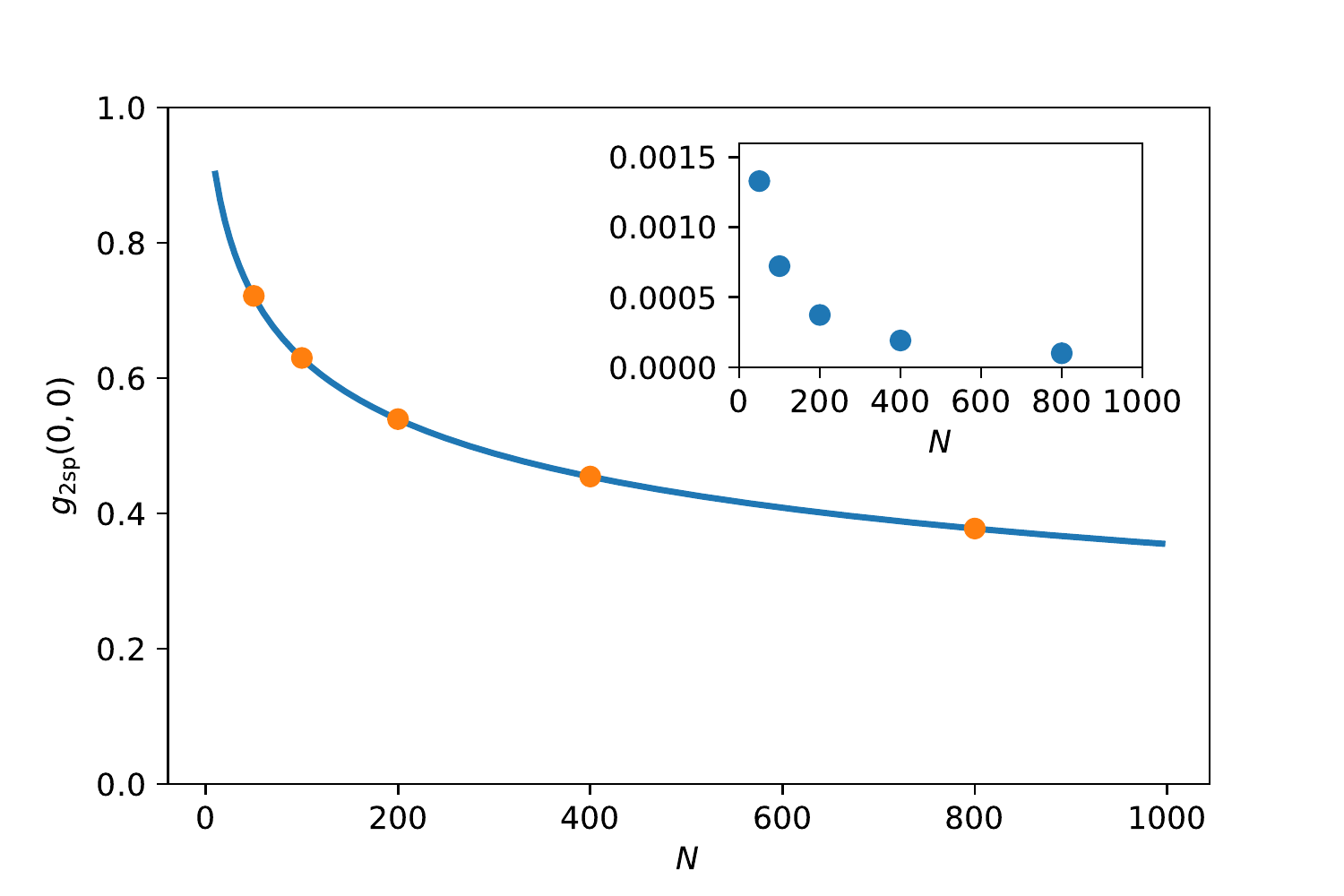}
	\caption{Sum rule $g_{2{\rm sp}}(0,0)$ from~\eqref{sum_rule_TL} (blue line) compared to the sum rule obtained by explicit summation for the finite system (orange, dots). In the inset we show the absolute value of the difference between the two results showing that they approach each other in the thermodynamic limit.}
	\label{fig:sum_rule}
\end{figure}

To this end we evaluate $g_{2{\rm sp}}(0,0)$. Using results of appendix~\ref{appB} we compute $J_0(0,0) = J_1(0,0) = 0$ and $J_2(0,0) = -1$. This gives $K(0,0) = -2$ and $L(0,0) = 0$. Therefore,
\begin{equation}
	g_{2{\rm sp}}(0,0) = \frac{\left( 2\pi n \right)^2 \Omega(L, N)}{\pi^4 N^2} \Xi_N \left( \Xi_N - 2\right). \label{sum_rule_TL}
\end{equation}
In Fig.~\ref{fig:sum_rule} we plot the sum rule as a function of the system size. The contribution from the $2{\rm sp}$ saturates the sum rule up to almost $50\%$ for $N=400$ and remains significant even for thousand particles.

\subsection*{Static correlator}

\begin{figure}
	\center
	\includegraphics[scale=0.5]{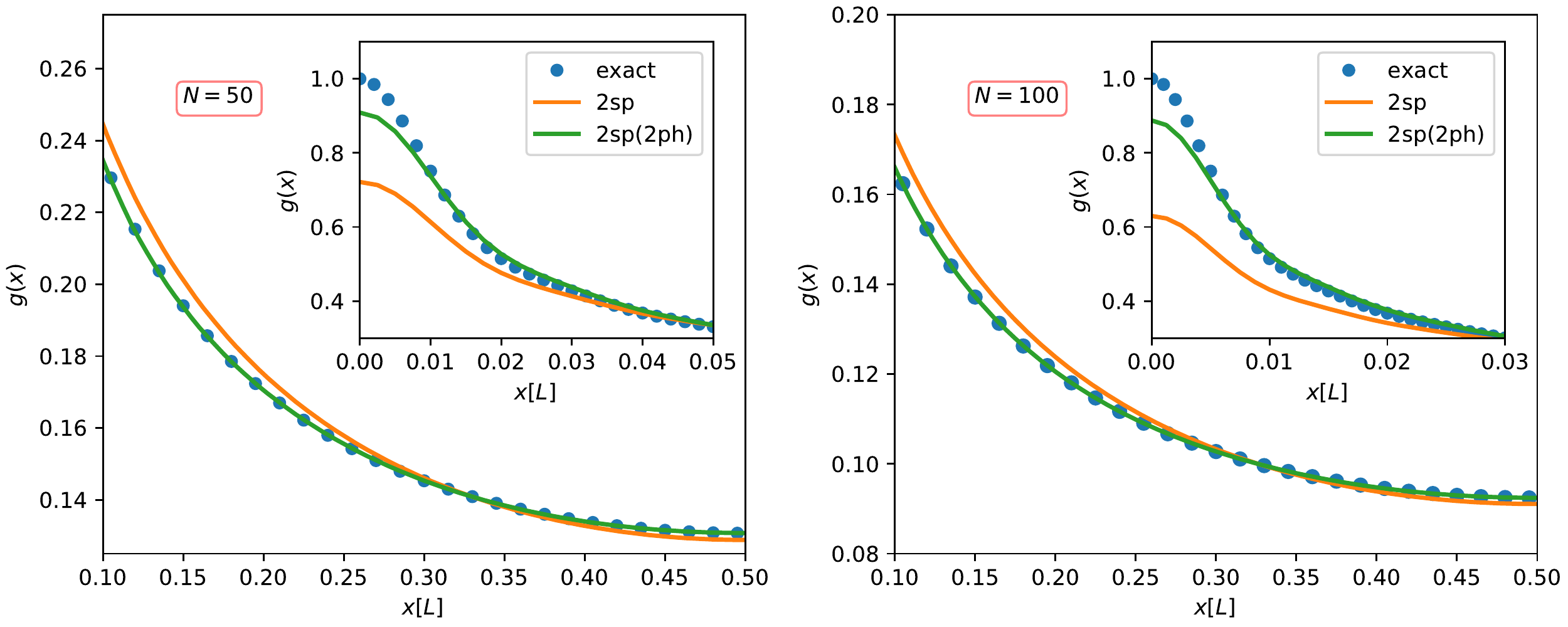}
	\caption{We plot the two-spinon contribution to the static one-body function (solid orange). Here the analytic result of Section~\ref{sec6} and the numeric result of Section~\ref{sec:TG} lie on top of each other. Therefore, we plot only the latter to make the figure clearer. We also show numeric result for dressed two-spinon excitations (solid green) and exact result by Lenard from~\eqref{leny} (blue points). The inset shows small $x$ behaviour. }
	\label{fig:SFT_TG_TL}
\end{figure}

We conclude our work by evaluating the equal-time correlator $g_{2{\rm sp}}(x, 0)$ which gives us an opportunity for direct comparison with Lenard's formula~\eqref{leny}. The special case of $t=0$ simplifies the computations because $d(x, 0; (1-z)) = - d(x, 0; z)$ and the following relations hold 
\begin{align}
	J_0(x, 0) = J_1(x, 0) + J_1^*(x, 0), \qquad J_2(x, 0) = J_1(x, 0) - \frac{\sin (k_F x)}{k_F x}.
\end{align}
Using that $d(x, 0;0) = -i k_F x$, we then find
\begin{align}
	K(x, 0) &= e^{-ik_F x} J_2^*(x,0) + e^{-i k_F x} J_2(x,0), \\
	L(x, 0) &= J_1(x,0) J_1^*(x,0) - J_0(x, 0) \frac{\sin (k_F x)}{k_F x},
\end{align} 
with both functions being evidently real. We can further manipulate these expressions by expressing $J_a(x,0)$ in terms of Sine and Cosine integrals, defined as 
\begin{align}
	{\rm Si}(x) = \int_0^x {\rm d}z \frac{\sin z}{z}, \qquad
	{\rm Ci}(x) = \int_x^{\infty} \frac{\cos z}{z}.
\end{align}
The Cosine integral has a logarithmic singularity at $x \rightarrow 0^+$. 
This singular behavior can be extracted by introducing 
\begin{equation}
	{\rm Cin}(x) = \int_0^x {\rm d} z\frac{1-\cos z}{z}.
\end{equation} 
These two functions are related by
\begin{equation}
	{\rm Ci}(x) = \gamma + \ln x - {\rm Cin}(x).
\end{equation}
On the other hand for $x > 0$ we have the following relation 
\begin{equation}
	\fint_0^x {\rm d}z \frac{\cos z}{z} = \fint_0^x {\rm d}z \frac{\cos z - 1}{z} + \fint_0^x  \frac{{\rm d}z}{z} = - {\rm Cin}(x),
\end{equation}
where in the second step we used that the first integral is a standard Lebesgue integral and the second integral vanishes because of the regularization. We then have 
\begin{align}
	J_1(x,0) &= \fint_0^1 {\rm d}z \frac{e^{id(x, 0;z\rho)}}{1-z} = \fint_0^1 {\rm d}z \frac{e^{-id(x, 0 ;z\rho)}}{z} = e^{i k_F x}\fint_0^1 {\rm d}z \frac{e^{-2i k_F x z}}{z} = e^{i k_F x} \fint_0^{2k_F x} {\rm d}z \frac{e^{-iz}}{z} \\
	&= e^{i k_F x} \fint_0^{2k_F x} {\rm d}z \frac{\cos z - i \sin z}{z} = - e^{i k_F x} \left( {\rm Cin}(2k_F x) - i\, {\rm Si}(2k_F x) \right).
\end{align}
This leads to 
\begin{align}
	K(x, 0) &= - 2\left( {\rm Cin}(2k_F x) + \frac{\sin(2k_F x)}{2k_F x} \right), \\
	L(x, 0) %&= {\rm Cin}^2(2k_F x)  + {\rm Si}^2(2k_F x) + {\rm Cin}(2k_F x)\frac{2\sin(2k_F x)}{2k_F x} - 2\frac{1 - \cos (2k_F x)}{2k_F x} {\rm Si}(2k_F x) \nonumber \\
	&= {\rm Cin}(2k_F x)\left( {\rm Cin}(2k_F x) + \frac{2\sin(2k_F x)}{2k_F x}\right) + {\rm Si}(2k_F x) \left({\rm Si}(2k_F x) - 2\frac{1 - \cos (2k_F x)}{2k_F x} \right),
\end{align}
and $g_{2{\rm sp}}(x,0)$ can be evaluated with~\eqref{final_g_2sp}. In fig.~\ref{fig:SFT_TG_TL} we plot $g_{2{\rm sp}}(x, 0)$ and compare it with Lenard's formula and with finite-system evaluation of the one-body function discussed in Section~\ref{sec:TG}. Results show that the bare two-spinon contribution describes qualitatively the full correlator. The largest discrepancy is for small distances. Instead, the dressed the two-spinon contribution matches the full correlator quantitatively over all distances but very short where large momentum excitations, not captured by our approximations are important.  

\subsection*{Asymptotic expansion and Luttinger liquid theory}
We conclude by evaluating the asymptotic expansion of $g_{2{\rm sp}}(x,0)$ for $x \gg k_F^{-1}$. This is the regime of the Luttinger liquid theory which predicts~\cite{1981_Haldane_PRL_47,Cazalilla_2004}
\begin{equation}
	g(x,0) \sim 1 - \frac{1}{(k_F x)^{1/2}} + \dots
\end{equation}
On the other hand, in the asymptotic expansion, $x \gg k_F^{-1}$, we have 
\begin{align}
	{\rm Si}(2k_F x) = \frac{\pi}{2} - \frac{\cos (2k_F x)}{2k_F x} + \dots , \qquad
	{\rm Ci}(2k_F x) = \frac{\sin 2k_F x}{2k_F x} + \dots.
\end{align}
Then, $K(x,0) = - 2 \gamma + \mathcal{O}((k_F x)^{-2})$ and $L(x,0) = \gamma^2 + \frac{\pi^2}{4} - \frac{\pi}{2k_F x} + \mathcal{O}((k_F x)^{-2})$ and consequently, for the correlation function, we find that it decays as $1/(k_F x)$. Therefore, the contribution from just two-spinon states is not enough to correctly predict the exponent of the power-law behaviour. We conjecture that dressing with the particle-hole excitations is required for that.

\section{Summary and outlook}\label{summary}

In this work we have studied the ground state one-body function of Lieb-Liniger model. Our approach was driven by the following question. Is there a well-defined class of excitations that is responsible for the small momentum-energy part of the correlation? We have argued that two-spinon states --- two hole excitations over the $(N+1)$-particle ground state --- together with particle-hole excitations organize the spectral sum in the desired way. 

We have verified this proposition with the help of the ABACUS for values of interaction strength $c$ ranging from weakly to strongly interacting case for $G(k,\omega)$ but also for integrated correlators $G(k)$ and $G(\omega)$. This analysis has also shown that from the point of view the spectral sum, the strongly interacting case, the Tonks-Girardeau gas is the most complex. On the other hand the Tonks-Girardeau gas allows us to pursue the analytic approach with relative simplicity. This allowed us to explicitly show two things. First, that the spectral sum can be organized such that the form factors of descendent states, containing more particle-hole excitations are monotonically decreasing. Second, that organizing the spectral sum around the two-spinon states is more effective than organizing it in terms of the particle-hole excitations over the $(N-1)$-particle ground state. We expect that, in view of the results at finite $c$, both observations hold for the Lieb-Liniger model as well. 

We concluded our work with computation of the two-spinon contribution in the large system, deriving an analytic expression for $g_{2{\rm sp}}(x,t)$. We have shown that this expression provides a relatively good approximation to the full correlation function. Specifically, saturates the sum-rule up to $1/3$ for systems containing as many as thousand particles. However, the fine structure of the $g(x,t)$ escapes this approximation. At short distances $g_{2{\rm sp}}(x,0)$ varies significantly from Lenard's formula. This should not be a surprise as our approach focuses on capturing the small momentum-energy part of the correlator. In that respect, the bare $2{\rm sp}$ states are also not enough for the asymptotic, large distance expansion of the correlation. The correlator decays as $1/(k_F x)$ instead of $1/\sqrt{k_F x}$ as predicted by the Luttinger liquid theory. This is understood, as recovering the Luttinger liquid requires soft-modes summation~\cite{1742-5468-2012-09-P09001}.

This work leads to a number of open questions. The most obvious one concerns generalization of the analytic expression for the two-spinon contribution to finite interaction strengths. Contrary to the Tonks-Girardeau case there is no easily computable equivalent of Lenard's formula in the Lieb-Liniger model. Therefore an explicit although approximate, formula would be welcomed. On a more technical level, an open question is how to analytically perform the soft-mode summation dressing the two-spinon excitations. Similar problems were recently successfully addressed in the Lieb-Liniger model~\cite{10.21468/SciPostPhys.9.6.082,Granet_2021} and in spin chains~\cite{10.21468/SciPostPhys.9.3.033,Gamayun_2021}. We hope to bring those techniques to this context. Our findings might also provide new ideas for optimization of the ABACUS scanning algorithms which would allow to access larger system sizes. 

The ultimate aim of our approach is to pursue computation of the one-body function on any finite energy density state, example being the finite temperature states or non-equilibrium stationary states. Parallel program for the density-density correlation function turned out to be quite successful and resulted in determining the thermodynamic form factors of the density operator. Our work should help in formulating thermodynamic form factors of the annihilation/creation operators or more generally of semi-local operators.

\section*{Acknowledgements} The authors are grateful to Jean-S{\'e}bastien Caux for granting them the access to the ABACUS data. The authors also acknowledge support from the National Science Centre, Poland, under the SONATA grant~2018/31/D/ST3/03588.

\appendix

\section{Form factors in the Tonks-Girardeau gas} \label{appA}

In this appendix, we work out the Tonks-Girardeau limit of the form factor \eqref{ff_LL}. 
Firstly, let us analyze the matrix $U(\bfm, \bfl)$ entries \eqref{eq.U} as $c\to \infty$.
From \eqref{eq.kernel}, we have $K(\lambda) = 2/c + \mathcal{O}\left(c^{-3}\right)$. This yields that 
\begin{equation}
\left[K(\lambda_{j}-\lambda_{k})-K(\lambda_{N}-\lambda_{k})\right]\sim \mathcal{O}\left(c^{-3}\right).
\end{equation}
Consequently, the leading contribution to the asymptotic behaviour of the matrix elements $U_{jk}(\bfm,\bfl)$ comes from $V_j^{\pm}$ defined in~\eqref{eq.Vj}. We find $V_j^\pm =  \mp i/c + \mathcal{O}\left(c^{-2} \right)$ and therefore
\begin{equation}
\left(\det_{N-1} U(\bfm, \bfl)\right)^2 = \left(\frac{2}{c}\right)^{2(N-1)} \times \left( 1 + \mathcal{O}(1/c)\right).
\end{equation}
Now, taking a look at the norms \eqref{eq.norm}, we have that the Gaudin matrix entries \eqref{eq.gaudin} yield, in the TG limit, 
\begin{equation}
\mathcal{G}_{ij}(\bfl_N) = \delta_{ij}\left(L+\frac{2(N-1)}{c}\right)-\frac{2}{c} + \mathcal{O}(1/c^3).
\end{equation}
We observe that the diagonal terms provide the leading contribution to the determinant, so that 
\begin{equation}
\det_{N}\mathcal{G}(\bfl_N) =  \left(L+\frac{2(N-1)}{c}\right)^{N}\times \left(1 + \mathcal{O}(1/c^2) \right) = L^{N} \left(1 + \mathcal{O}(1/c) \right).
\end{equation}
Thence, 
\begin{equation}
\| \bfl_N\|^{2} =  L^{N} c^{N^2}\prod_{j>k=1}^{N}\lambda_{jk}^{-2} \times \left( 1 + \mathcal{O}(1/c) \right).
\end{equation}
%Likewise, 
%\begin{equation}
%\| \bfm_{N-1}\|^{2}\sim (Lc)^{N-1}\prod_{j>k=1}^{N-1}\frac{\mu_{jk}^{2}+c^{2}}{\mu_{jk}^{2}}.
%\end{equation}
Inserting all terms together, we have that the form factor in the TG limit results in 
\begin{align}
 |\langle \bfm | \Psi(0) | \bfl \rangle|^2 = \frac{1}{2} \left(\frac{2}{L}\right)^{2N-1}\frac{\prod_{j>k=1}^{N}\lambda_{jk}^{2}\prod_{j>k=1}^{N-1}\mu_{jk}^{2}}{\prod_{j=1}^{N}\prod_{k=1}^{N-1}(\lambda_{j}-\mu_{k})^{2}}.
\end{align}
%\begin{align}
% |\langle \bfm | \Psi(0) | \bfl \rangle|^2 
% & \sim \frac{1}{2} \left(\frac{2}{L}\right)^{2N-1}\ c^{-2(N-1)} \frac{\prod_{j>k=1}^{N}(\lambda_{jk}^{2}+c^{2})}{\prod_{j=1}^{N}\prod_{k=1}^{N-1}(\lambda_{j}-\mu_{k})^{2}}\frac{\prod_{j>k=1}^{N}\lambda_{jk}^{2}\prod_{j>k=1}^{N-1}\mu_{jk}^{2}}{\prod_{j>k=1}^{N-1}\left(\mu_{jk}^{2}+c^{2}\right)} \nonumber \\
% & = \frac{1}{2} \left(\frac{2}{L}\right)^{2N-1}\frac{\prod_{j>k=1}^{N}\lambda_{jk}^{2}\prod_{j>k=1}^{N-1}\mu_{jk}^{2}}{\prod_{j=1}^{N}\prod_{k=1}^{N-1}(\lambda_{j}-\mu_{k})^{2}}.
%\end{align}

\subsection{Prefactor}\label{subsec.prefactor}
We compute the excitation independent prefactor $\Omega(L, N)$ defined in~\eqref{OmegaNL}. We insert the ground state rapidities 
\begin{equation}
	\lambda_j = \frac{2\pi}{L}\left(- \frac{N+1}{2} + j\right), \qquad \bar{\mu}_k =\frac{2\pi}{L}\left( - \frac{N+2}{2} + k\right), 
\end{equation}
in $\Omega(N,L)$ to find
\begin{equation}
	\Omega(N, L) = \frac{L}{4} \left[ \pi^{-N} \frac{\prod_{j > k =1}^N (j - k)  \prod_{j > k =1}^{N+1} (j - k)}{\prod_{j=1}^N \prod_{k=1}^{N+1} \left(j - k + 1/2\right)} \right]^2.
\end{equation} 
The expression in the square bracket can be expressed in terms of the $\Gamma(z)$ function and Barnes $G(z)$ function. We have
\begin{equation}
	\prod_{j > k =1}^N (j - k) = \prod_{j=1}^N \prod_{k=1}^{j-1} (j-k) = \prod_{j=1}^N \prod_{k=1}^{j-1} k = \prod_{j=1}^N \Gamma(j) = G(N+1). 
\end{equation}
The product in the denominator can be rewritten as 
\begin{align}
	\prod_{j=1}^N \prod_{k=1}^{N+1} \left(j - k + 1/2\right) &= (-1)^{N(N+1)/2} \left(\prod_{j=1}^N \prod_{k=1}^j (k-1/2) \right)^2 \nonumber \\
	&=  (-1)^{N(N+1)/2}\left(\prod_{j=1}^N \frac{\Gamma(j+1/2)}{\Gamma(1/2)} \right)^2 \nonumber \\
	&= \frac{(-1)^{N(N+1)/2}}{\pi^N} \left(\frac{G(N+3/2)}{G(3/2)}\right)^2,
\end{align} 
where we used that $\Gamma(1/2) = \sqrt{\pi}$.
Therefore
\begin{equation}
	\Omega(N, L) = \frac{L}{4} G^4(3/2) \left[ \frac{G(N+1) G(N+2)}{G^2(N+3/2)} \right]^2.
\end{equation}
In the large $N$ limit the square bracket simplifies to 
\begin{equation}
	 \frac{G(N+1) G(N+2)}{G^2(N+3/2)} = N^{1/4} \times \left( 1 + \frac{1}{8N} + \mathcal{O}\left(N^{-2}\right)\right),
\end{equation}
and
\begin{equation}
	\Omega(N, L) = G^4(3/2) \frac{L N^{1/2}}{4} \times \left( 1 + \frac{1}{4N} + \mathcal{O}\left(N^{-2}\right)\right).
\end{equation}

\subsection{$m$-ph over the $(N-1)$-particle GS}
For the numerical evaluation of the one-body function in the Tonks-Girardeau we will also need form factors between the $N$-particle ground state and $m$ particle-hole excited state over the $(N-1)$-particle ground state.
Similarly to the derivation in Section~\ref{subsec:TG_ff}, let us consider now $m$ particle-hole excitations over the $(N-1)$-particle ground state characterized by the set of rapidities $\underline{\bfm}$. Then, the set of rapidities $\bfm$ of the $(N-1)$-particle excited state is 
\begin{equation}
	\bfm = \underline{\bfm} - \bfh_m + \bfp_m,
\end{equation}
The product over rapidities $\bfm$ reads 
\begin{equation}
	\prod_{j=1}^{N-1} f(\mu_j) = \prod_{j=1}^{N-1} f(\umu_j) \times \prod_{a=1}^{m} \frac{f(p_a)}{f(h_a)}.
\end{equation}
Then, the analog for the double products yields 
\begin{equation}
	\prod_{j \neq k} f(\mu_j, \mu_k)  =\prod_{j\neq k}f(\umu_{j},\umu_{k})\times\Dprod_{j,a}\frac{f(\umu_{j},p_{a})f(p_{a},\umu_{k})}{f(\umu_{j},h_{a})f(h_{a},\umu_{k})}\times\Dprod_{a,b}\frac{f(p_{a},p_{b})}{f(p_{a},h_{b})}\frac{f(h_{a},h_{b})}{f(h_{a},p_{b})}.
\end{equation}
Consequently, the form factor \eqref{ff_TG} becomes
\begin{align}
	|\langle \bfm | \Psi(0) | \bfl \rangle|^2 &= \tilde{\Omega}(L, N) \times \prod_{j=1}^{N}\prod_{a=1}^{m}\frac{(\lambda_{j}-p_{a})^{2}}{(\lambda_{j}-h_{a})^{2}}\times\Dprod_{j,a}\frac{(\umu_{j}-p_{a})^{2}}{(\umu_{j}-h_{a})^{2}}\times\Dprod_{a,b}\frac{(p_{a}-p_{b})}{(p_{a}-h_{b})}\frac{(h_{a}-h_{b})}{(h_{a}-p_{b})}, 
\end{align}
where 
\begin{equation}
\tilde{\Omega}(L, N) = \frac{1}{2}\left(\frac{2}{L}\right)^{2N-1}\frac{ \prod_{j>k=1}^{N}\lambda_{jk}^2 \prod_{j> k=1}^{N-1}\umu_{jk}^2}{\prod_{j=1}^{N}\prod_{k=1}^{N-1}(\lambda_{j}-\umu_{k})^{2}} = \left(\frac{2}{L}\right)^2 \Omega(L, N-1), 
\end{equation}
is the excited-state-independent factor.

\section{Sums over $G_N(k)$ as integrals}\label{appB}

In this appendix we compute the continuum approximation to the sum
\begin{equation}
	\pi^2 N \sum_{k=1}^{N+1} G_N(k) f(k/N),
\end{equation}
appearing in Section~\ref{sec6}. As noted there, function $G_N(k)$, in large $N$ limit diverges for $k$ close to $1$ and $N$. To control this divergence we introduce $k^*$, such that $N \gg k^* \gg 1$. We will use it to quantify the distance from the edges. We divide now the sum into three pieces
\begin{equation}
	\sum_{k=1}^{N+1} G_N(k) f(k/N) = \sum_{k=1}^{k^*} G_N(k) f(k/N) + \sum_{k=k^* + 1}^{N - k^* - 1 } G_N(k) f(k/N) + \sum_{k= N - k^*}^{N+1} G_N(k) f(k/L). 
\end{equation}
We start with the middle sum involving summation far from the edges. Using Euler-Maclaurin formula we find
\begin{equation}
	\sum_{k=k^*+1}^{N - k^*-1} G_N(k) f(k/N) = \frac{1}{\pi^2 N} \int_{x^*}^{1-x^*} {\rm d}x \frac{f(x)}{x(1-x)} + \mathcal{O}(1/N^2),
\end{equation}
with $x^* \equiv k^*/ (N+1)$. Here we used the first term of the asymptotic expansion for the ratio of the $\Gamma(x)$ functions
\begin{equation}
	\frac{\Gamma(x-1/2)}{\Gamma(x)} = \frac{1}{x^{1/2}} + \frac{3}{8} \frac{1}{x^3/2} + \mathcal{O}\left(\frac{1}{x^{5/2}} \right),
\end{equation}
which leads to
\begin{equation}
	G_N(x (N+1)) \sim \frac{1}{\pi^2 N^2} \frac{1}{x(1-x)}, \quad N \rightarrow \infty.
\end{equation}
The integral can be rewritten as the standard principal value integral plus a boundary term depending on $x^*$. 
To this end, we introduce a small parameter $\epsilon >0$. Then, 
\begin{align}
	\int_{x^*}^{1-x^*} {\rm dx} \frac{f(x)}{x(1-x)} &= \int_{\epsilon}^{1 - \epsilon} {\rm d}x \frac{f(x)}{x(1-x)} - \int_{\epsilon}^{x^*} {\rm d}x \frac{f(x)}{x(1-x)} - \int_{1-x^*}^{1-\epsilon} {\rm d}x \frac{f(x)}{x(1-x)} \nonumber \\
	&= \int_{\epsilon}^{1 - \epsilon} {\rm d}x \frac{f(x)}{x(1-x)} -  f(0)\int_{\epsilon}^{x^*} \frac{{\rm d}x}{x} - f(1)\int_{1-x^*}^{1-\epsilon}  \frac{{\rm d}x}{(1-x)} + \mathcal{O}(x^*) \nonumber \\
	&= \int_{\epsilon}^{1 - \epsilon} {\rm d}x \frac{f(x)}{x(1-x)} + (f(0) + f(1)) \ln \epsilon - (f(0) + f(1)) \ln x^*.
\end{align}
The first two terms have a finite value in the $\epsilon\rightarrow 0$ limit given by the principal value integral defined as
\begin{equation}
	\fint_0^1 {\rm d}x \frac{f(x)}{x(1-x)} \equiv \lim_{\epsilon \rightarrow 0} \left( \int_{\epsilon}^{1 - \epsilon} {\rm d}x \frac{f(x)}{x(1-x)} + (f(0) + f(1)) \ln \epsilon\right).\label{pvint}
\end{equation}
Therefore
\begin{equation}
	\int_{x^*}^{1-x^*} {\rm dx} \frac{f(x)}{x(1-x)} = \fint_0^1 {\rm d}x \frac{f(x)}{x(1-x)} - (f(0) + f(1)) \ln x^* + \mathcal{O}(x^*),
\end{equation}
and
\begin{equation}
	\sum_{k=k^*+1}^{N - k^*-1} G_N(k) f(k/N) = \frac{1}{\pi^2 N} \left(\fint_0^1 {\rm d}x \frac{f(x \rho)}{x(1-x)} - (f(0) + f(1) \ln (k^*/N) + \mathcal{O}(k^*/N) \right). 
\end{equation}
This is the final result for the middle sum.

Now, let us consider one of the edge sums. Provided that $k^* \ll N$, we have  
\begin{align}
	\sum_{k=1}^{k^*} G_N(k) f(k/N) &= \sum_{k=1}^{k^*} G_N(k) ( f(1/N) + \mathcal{O}(k^*/N) ) \approx f(1/N) \sum_{k=1}^{k^*} G_N(k).
\end{align}
We are interested in the behaviour of the sum at large $N$ and $k^*$ such that $N - k^* \gg 1$. Therefore, we can approximate the second part of $G_N(k)$ using
\begin{equation}
	\left( \frac{\Gamma(x + 3/2}{\Gamma(x + 2)}\right)^2 = \frac{1}{x} - \frac{5}{4x^2} + \mathcal{O}\left(x^{-3}\right),
\end{equation} 
yielding 
\begin{equation}
	\sum_{k=1}^{k^*} G_N(k) = \frac{1}{\pi^2}\sum_{k=1}^{k^*} \left(\frac{\Gamma(k - 1/2}{\Gamma(k)}\right)^2 \times \left( \frac{1}{N-k} - \frac{5}{4(N-k)^2} + \mathcal{O}\left((N-k)^{-3}\right)\right).
\end{equation}
We want the result with the same precision as for the bulk part, $\mathcal{O}\left(k^*/N^2\right)$. Therefore, we further expand the bracket using $k^* \ll N$ to find that
\begin{equation}
	\frac{1}{N-k} - \frac{5}{4(N-k)^2} \approx \frac{1}{N} \left(1 + \frac{k}{N} + \mathcal{O}\left((k^*/N)^2\right)\right).
\end{equation}
Consequently,
\begin{equation}
	\sum_{k=1}^{k^*} G_N(k) = \frac{1}{\pi^2 N}\sum_{k=1}^{k^*} \left(\frac{\Gamma(k - 1/2}{\Gamma(k)}\right)^2 \times \left(1 + \frac{k}{N} + \mathcal{O}\left((k^*/N)^2\right)\right).
\end{equation}
To proceed further, we analyze the two sums separately. Regarding the first one, we rewrite
\begin{equation}
	\sum_{k=1}^{k^*} \left(\frac{\Gamma(k - 1/2}{\Gamma(k)}\right)^2 = \sum_{k=1}^{k^*} \left( \left(\frac{\Gamma(k - 1/2}{\Gamma(k)}\right)^2 - \frac{1}{k}\right) + H_{k^*},
\end{equation}
where $H_n$ is the $n$-th Harmonic number defined by  
\begin{equation}
	H_n \equiv \sum_{k=1}^n \frac{1}{k}.
\end{equation}
The remaining sum converges to 
\begin{equation}
	A \equiv \lim_{k^*\rightarrow \infty}\sum_{k=1}^{k^*} \left( \left(\frac{\Gamma(k - 1/2}{\Gamma(k)}\right)^2 - \frac{1}{k}\right) \approx 2.77221,
\end{equation}
as we estimated numerically using Shanks transformation to speed up the convergence
For the Harmonic number we use the relation $H_n = \ln n + \gamma + \mathcal{O}(1/n)$, where $\gamma$ is the Euler-Mascheroni constant. 
Therefore, 
\begin{equation}
	\sum_{k=1}^{k^*} \left(\frac{\Gamma(k - 1/2}{\Gamma(k)}\right)^2 = \ln k^* + A + \gamma + \mathcal{O}(1/k^*).
\end{equation}
The other sum can be computed in the following way
\begin{equation}
	\sum_{k=1}^{k^*} \left(\frac{\Gamma(k - 1/2}{\Gamma(k)}\right)^2 \frac{k}{N} = \frac{k^*}{N} \sum_{k=1}^{k^*} \left(\frac{\Gamma(k - 1/2}{\Gamma(k)}\right)^2 \frac{k}{k^*} = \frac{k^*}{N} C, 
\end{equation}
where 
\begin{equation}
	C \equiv \lim_{k^*\rightarrow \infty} \sum_{k=1}^{k^*} \left(\frac{\Gamma(k - 1/2}{\Gamma(k)}\right)^2 \frac{k}{k^*} \approx 1.00015,
\end{equation}
was again estimated numerically with the help of Shanks transformation. The term proportional to $C$ is subleading in $k^*/N \ll 1$ limit and we find the following expression for the edge sum
\begin{equation}
	\sum_{k=1}^{k^*} G_N(k) = \frac{1}{\pi^2 N} \left( \ln k^* + A + \gamma + \frac{k^*}{N} C \right).
\end{equation}
The other boundary sum gives similar expression and we find that 
\begin{equation}
	\sum_{k=1}^{N+1} G_N(k) f(k/N) = \frac{1}{\pi^2 N}\fint_0^1 {\rm d}x \frac{f(x)}{x(1-x)} + \frac{1}{\pi^2 N} \left( f(0) + f(1) \right) \left(\ln k^* - \ln x^* + A + \gamma \right) + \mathcal{O}(k^*/N^2).
\end{equation}
Making use of $x^* = k^* / (N+1)$, we find
\begin{equation}
	\sum_{k=1}^{N+1} G_N(k) f(k/N) = \frac{1}{\pi^2 N}\fint_0^1 {\rm d}x \frac{f(x )}{x(1-x)} + \frac{f(0) + f(1)}{\pi^2 N}  \left(\ln (N+1) + A + \gamma \right) + \mathcal{O}(k^*/N^2). \label{app_single_sum}
\end{equation}
This is the final expression that transforms a sum weighted with $G_N(k)$ into an integral plus boundary terms. We note that the boundary terms, relative to the integral, are logarithmically diverging in the large $N$ limit.

\subsection*{Examples}
We apply now formula~\eqref{single_sum} for some simple functions $f(x)$. %Then, we check that~\eqref{single_sum} is correct up to $\mathcal{O}\left(N^{-2}\right)$.  
We consider $f(x) = 1,x,x^2$. The results are 
\begin{align}
	\sum_{k=1}^{N+1} G_N(k) &= \frac{2}{\pi^2 N}  \left(\ln (N+1) + A + \gamma \right), \label{ex1} \\
	\sum_{k=1}^{N+1} G_N(k) (k/N) &= \frac{1}{\pi^2 N}  \left(\ln (N+1) + A + \gamma \right), \label{ex2} \\
	\sum_{k=1}^{N+1} G_N(k) (k/N)^2 &= -\frac{1}{\pi^2 N} + \frac{1}{\pi^2 N}  \left(\ln (N+1) + A + \gamma \right),  \label{ex3}
\end{align}
where we used that  
\begin{equation}
	\fint_0^1  \frac{{\rm d}x}{x(1-x)} = 0, \qquad \fint_0^1  {\rm d}x\frac{x}{x(1-x)} = 0, \qquad \fint_0^1  {\rm d}x\frac{x^2}{x(1-x)} = -1.
\end{equation}
In all these expressions there are subleading terms of order $N^{-2}$ or higher that were neglected. The results of this section are used in Section~\ref{sec6} when evaluating the sum-rule. 

\bibliographystyle{JHEP}
\bibliography{biblio}

\end{document}